\documentclass[journal=jacsat,manuscript=article]{achemso}

\usepackage[version=3]{mhchem} % Formula subscripts using \ce{}
\usepackage{xcolor}
\usepackage{bm}
\usepackage{tabularx}
\usepackage{siunitx}
\usepackage{makecell}
\usepackage{tikz}
\author{Tianyu Fang}
\affiliation[IAP]
{Institute of Applied Physics, University of Bonn, Wegelerstr. 8, Bonn, Germany}

\author{Ricardo Gioia Alvarez}
\affiliation[IAP]
{Institute of Applied Physics, University of Bonn, Wegelerstr. 8, Bonn, Germany}

\author{Babak Behjati}
\affiliation[IAP]
{Institute of Applied Physics, University of Bonn, Wegelerstr. 8, Bonn, Germany}

\author{Moritz Scharfst{\"a}dt}
\affiliation[PI]
{Physikalisches Institut, University of Bonn, Nussallee 12, Bonn, Germany}

\author{{Max Masuhr}}
\affiliation[IAP]
{Institute of Applied Physics, University of Bonn, Wegelerstr. 8, Bonn, Germany}

\author{{Bo Deng}}
\affiliation[IAP]
{Institute of Applied Physics, University of Bonn, Wegelerstr. 8, Bonn, Germany}

\author{Noah Henseler}
\affiliation[IAP]
{Institute of Applied Physics, University of Bonn, Wegelerstr. 8, Bonn, Germany}

\author{Andrea Bergschneider}
\affiliation[PI]
{Physikalisches Institut, University of Bonn, Nussallee 12, Bonn, Germany}

\author{Stefan Linden}
\affiliation[PI]
{Physikalisches Institut, University of Bonn, Nussallee 12, Bonn, Germany}

\author{Christian Sch{\"a}fer}
\affiliation[Wien]
{Institute of Applied Physics, Technical University of Vienna, Wiedner Hauptstr. 8-10, Vienna, Austria}

\author{Daqing Wang}
\affiliation[IAP]
{Institute of Applied Physics, University of Bonn, Wegelerstr. 8, Bonn, Germany}
\email{daqing.wang@uni-bonn.de}

\title{Aromatic molecular emitters in a hexagonal boron nitride stack}

\abbreviations{IR,NMR,UV}

\usepackage[T1]{fontenc}
\usepackage{amsmath}
\usepackage{enumitem}
\usepackage{titlesec}

\newcommand{\CombinedSITitlePage}{%
  \thispagestyle{empty}%
  \begin{center}%
  {\LARGE\bfseries Supplementary Information for\\[0.5em]%
  Aromatic molecular emitters in a hexagonal boron nitride stack\par}%
  \vspace{1.75em}%
  {\large Tianyu Fang,\textsuperscript{1} Ricardo Gioia Alvarez,\textsuperscript{1} Babak Behjati,\textsuperscript{1} Moritz Scharfst{\"a}dt,\textsuperscript{2} Max Masuhr,\textsuperscript{1} Bo Deng,\textsuperscript{1} Noah Henseler,\textsuperscript{1} Andrea Bergschneider,\textsuperscript{2} Stefan Linden,\textsuperscript{2} Christian Sch{\"a}fer,\textsuperscript{3} Daqing Wang\textsuperscript{1,*}\par}%
  \vspace{1.5em}%
  {\normalsize \textsuperscript{1}Institute of Applied Physics, University of Bonn, Wegelerstr. 8, Bonn, Germany\par}%
  \vspace{0.5em}%
  {\normalsize \textsuperscript{2}Physikalisches Institut, University of Bonn, Nussallee 12, Bonn, Germany\par}%
  \vspace{0.5em}%
  {\normalsize \textsuperscript{3}Institute of Applied Physics, Technical University of Vienna, Wiedner Hauptstr. 8-10, Vienna, Austria\par}%
  \vspace{1em}%
  {\normalsize \textsuperscript{*}daqing.wang@uni-bonn.de\par}%
  \end{center}%
}

\newcommand{\CombinedSITableOfContents}{%
  \section*{Contents}%
  \contentsline {section}{\numberline {1}Sample preparation}{1}{}%
  \contentsline {subsection}{\numberline {1.1}Evaporation of perylene molecules}{1}{}%
  \contentsline {subsection}{\numberline {1.2}Fabrication of hBN stacks}{3}{}%
  \contentsline {subsection}{\numberline {1.3}Fluorescence images}{7}{}%
  \contentsline {subsection}{\numberline {1.4}Perylene in anthracene sample fabrication}{8}{}%
  \contentsline {section}{\numberline {2}Experimental setup}{9}{}%
  \contentsline {section}{\numberline {3}Numerical calculations}{11}{}%
  \contentsline {subsection}{\numberline {3.1}Visualization of insertion geometries}{11}{}%
  \contentsline {subsection}{\numberline {3.2}Diffusion barrier of perylene in an hBN stack}{17}{}%
  \contentsline {section}{\numberline {4}Emission spectra}{18}{}%
  \contentsline {subsection}{\numberline {4.1}Calculation of Franck-Condon factors}{18}{}%
  \contentsline {subsection}{\numberline {4.2}Phonon broadening model}{20}{}%
  \contentsline {subsection}{\numberline {4.3}Temperature-dependent emission spectra}{21}{}%
  \contentsline {section}{\numberline {5}Spectral diffusion data analysis}{23}{}%
  \contentsline {section}{\numberline {6}Sample Annealing}{27}{}%
  \contentsline {section}{\numberline {7}Hyperspectral imaging}{28}{}%
  \contentsline {subsection}{\numberline {7.1}LNCC matching}{30}{}%
  \contentsline {subsection}{\numberline {7.2}Peak-spacing matching}{31}{}%
}

\newcommand{\StartCombinedSI}{%
  \clearpage%
  \CombinedSITitlePage%
  \clearpage%
  {\pagestyle{empty}\CombinedSITableOfContents\clearpage}%
  \setcounter{page}{1}%
  \setcounter{section}{0}%
  \setcounter{subsection}{0}%
  \setcounter{figure}{0}%
  \setcounter{table}{0}%
  \setcounter{equation}{0}%
  \renewcommand{\thesection}{\arabic{section}}%
  \renewcommand{\thefigure}{\textbf{S\arabic{figure}}}%
  \renewcommand{\figurename}{\textbf{Figure}}%
  \renewcommand{\thetable}{\textbf{S\arabic{table}}}%
  \renewcommand{\tablename}{\textbf{Table}}%
  \renewcommand{\theequation}{S\arabic{equation}}%
  \providecommand*{\theHpage}{\arabic{page}}%
  \providecommand*{\theHsection}{\thesection}%
  \providecommand*{\theHsubsection}{\thesubsection}%
  \providecommand*{\theHfigure}{\thefigure}%
  \providecommand*{\theHtable}{\thetable}%
  \providecommand*{\theHequation}{\theequation}%
  \renewcommand{\theHpage}{SI.\arabic{page}}%
  \renewcommand{\theHsection}{SI.\arabic{section}}%
  \renewcommand{\theHsubsection}{SI.\arabic{section}.\arabic{subsection}}%
  \renewcommand{\theHfigure}{SI.\arabic{figure}}%
  \renewcommand{\theHtable}{SI.\arabic{table}}%
  \renewcommand{\theHequation}{SI.\arabic{equation}}%
}
\makeatletter
\gdef\CombinedMainBibliography{%
\providecommand{\latin}[1]{##1}
\makeatletter
\providecommand{\doi}
  {\begingroup\let\do\@makeother\dospecials
  \catcode`\{=1 \catcode`\}=2 \doi@aux}
\providecommand{\doi@aux}[1]{\endgroup\texttt{##1}}
\makeatother
\providecommand*\mcitethebibliography{\thebibliography}
\csname @ifundefined\endcsname{endmcitethebibliography}
  {\let\endmcitethebibliography\endthebibliography}{}

}

\gdef\CombinedSIBibliography{%
\providecommand{\latin}[1]{##1}
\makeatletter
\providecommand{\doi}
  {\begingroup\let\do\@makeother\dospecials
  \catcode`\{=1 \catcode`\}=2 \doi@aux}
\providecommand{\doi@aux}[1]{\endgroup\texttt{##1}}
\makeatother
\providecommand*\mcitethebibliography{\thebibliography}
\csname @ifundefined\endcsname{endmcitethebibliography}
  {\let\endmcitethebibliography\endthebibliography}{}
\begin{mcitethebibliography}{22}
\providecommand*\natexlab[1]{##1}
\providecommand*\mciteSetBstSublistMode[1]{}
\providecommand*\mciteSetBstMaxWidthForm[2]{}
\providecommand*\mciteBstWouldAddEndPuncttrue
  {\def\EndOfBibitem{\unskip.}}
\providecommand*\mciteBstWouldAddEndPunctfalse
  {\let\EndOfBibitem\relax}
\providecommand*\mciteSetBstMidEndSepPunct[3]{}
\providecommand*\mciteSetBstSublistLabelBeginEnd[3]{}
\providecommand*\EndOfBibitem{}
\mciteSetBstSublistMode{f}
\mciteSetBstMaxWidthForm{subitem}{(\alph{mcitesubitemcount})}
\mciteSetBstSublistLabelBeginEnd
  {\mcitemaxwidthsubitemform\space}
  {\relax}
  {\relax}

\bibitem[Novoselov \latin{et~al.}(2004)Novoselov, Geim, Morozov, Jiang, Zhang,
  Dubonos, Grigorieva, and Firsov]{Geim_2004}
Novoselov,~K.~S.; Geim,~A.~K.; Morozov,~S.~V.; Jiang,~D.; Zhang,~Y.;
  Dubonos,~S.~V.; Grigorieva,~I.~V.; Firsov,~A.~A. Electric field effect in
  atomically thin carbon films. \emph{Science} \textbf{2004}, \emph{306},
  666--669\relax
\mciteBstWouldAddEndPuncttrue
\mciteSetBstMidEndSepPunct{\mcitedefaultmidpunct}
{\mcitedefaultendpunct}{\mcitedefaultseppunct}\relax
\EndOfBibitem
\bibitem[Neese(2025)]{SI-neese2025}
Neese,~F. Software Update: The ORCA Program System—Version 6.0. \emph{WIREs
  Computational Molecular Science} \textbf{2025}, \emph{15}, e70019\relax
\mciteBstWouldAddEndPuncttrue
\mciteSetBstMidEndSepPunct{\mcitedefaultmidpunct}
{\mcitedefaultendpunct}{\mcitedefaultseppunct}\relax
\EndOfBibitem
\bibitem[Lee \latin{et~al.}(1988)Lee, Yang, and Parr]{SI-lee1988development}
Lee,~C.; Yang,~W.; Parr,~R.~G. Development of the Colle-Salvetti
  correlation-energy formula into a functional of the electron density.
  \emph{Phys. Rev. B} \textbf{1988}, \emph{37}, 785\relax
\mciteBstWouldAddEndPuncttrue
\mciteSetBstMidEndSepPunct{\mcitedefaultmidpunct}
{\mcitedefaultendpunct}{\mcitedefaultseppunct}\relax
\EndOfBibitem
\bibitem[Becke(1993)]{SI-becke1993density}
Becke,~A.~D. Density-functional thermochemistry. III. The role of exact
  exchange. \emph{The Journal of Chemical Physics} \textbf{1993}, \emph{98},
  5648--5652\relax
\mciteBstWouldAddEndPuncttrue
\mciteSetBstMidEndSepPunct{\mcitedefaultmidpunct}
{\mcitedefaultendpunct}{\mcitedefaultseppunct}\relax
\EndOfBibitem
\bibitem[Caldeweyher \latin{et~al.}(2019)Caldeweyher, Ehlert, Hansen,
  Neugebauer, Spicher, Bannwarth, and Grimme]{Caldeweyher2019}
Caldeweyher,~E.; Ehlert,~S.; Hansen,~A.; Neugebauer,~H.; Spicher,~S.;
  Bannwarth,~C.; Grimme,~S. A generally applicable atomic-charge dependent
  London dispersion correction. \emph{The Journal of Chemical Physics}
  \textbf{2019}, \emph{150}, 154122\relax
\mciteBstWouldAddEndPuncttrue
\mciteSetBstMidEndSepPunct{\mcitedefaultmidpunct}
{\mcitedefaultendpunct}{\mcitedefaultseppunct}\relax
\EndOfBibitem
\bibitem[Casida(1996)]{SI-casida1996time}
Casida,~M.~E. Time-dependent density functional response theory of molecular
  systems: theory, computational methods, and functionals. \emph{Theoretical
  and Computational Chemistry} \textbf{1996}, \emph{4}, 391--439\relax
\mciteBstWouldAddEndPuncttrue
\mciteSetBstMidEndSepPunct{\mcitedefaultmidpunct}
{\mcitedefaultendpunct}{\mcitedefaultseppunct}\relax
\EndOfBibitem
\bibitem[Hjorth~Larsen \latin{et~al.}(2017)Hjorth~Larsen, Mortensen, Blomqvist,
  Castelli, Christensen, Du{\l}ak, Friis, Groves, Hammer, Hargus, Hermes,
  Jennings, Jensen, Kermode, Kitchin, Kolsbjerg, Kubal, Kaasbjerg, Lysgaard,
  Maronsson, Maxson, Olsen, Pastewka, Peterson, Rostgaard, Schi{\o}tz,
  Sch{\"u}tt, Strange, Thygesen, Vegge, Vilhelmsen, Walter, Zeng, and
  Jacobsen]{SI-hjorth2017atomic}
Hjorth~Larsen,~A. \latin{et~al.}  The atomic simulation environment—a Python
  library for working with atoms. \emph{Journal of Physics: Condensed Matter}
  \textbf{2017}, \emph{29}, 273002\relax
\mciteBstWouldAddEndPuncttrue
\mciteSetBstMidEndSepPunct{\mcitedefaultmidpunct}
{\mcitedefaultendpunct}{\mcitedefaultseppunct}\relax
\EndOfBibitem
\bibitem[Batatia \latin{et~al.}(2025)Batatia, Benner, Chiang, Elena, Kovács,
  Riebesell, Advincula, Asta, Avaylon, Baldwin, Berger, Bernstein, Bhowmik,
  Bigi, Blau, Cărare, Ceriotti, Chong, Darby, De, Della~Pia, Deringer,
  Elijošius, El-Machachi, Fako, Falcioni, Ferrari, Gardner, Gawkowski,
  Genreith-Schriever, George, Goodall, Grandel, Grey, Grigorev, Han, Handley,
  Heenen, Hermansson, Ho, Hofmann, Holm, Jaafar, Jakob, Jung, Kapil, Kaplan,
  Karimitari, Kermode, Kourtis, Kroupa, Kullgren, Kuner, Kuryla, Liepuoniute,
  Lin, Margraf, Magdău, Michaelides, Moore, Naik, Niblett, Norwood, O’Neill,
  Ortner, Persson, Reuter, Rosen, Rosset, Schaaf, Schran, Shi, Sivonxay,
  Stenczel, Sutton, Svahn, Swinburne, Tilly, van~der Oord, Vargas,
  Varga-Umbrich, Vegge, Vondrák, Wang, Witt, Wolf, Zills, and
  Csányi]{batatia2023foundation2}
Batatia,~I. \latin{et~al.}  A foundation model for atomistic materials
  chemistry. \emph{The Journal of Chemical Physics} \textbf{2025}, \emph{163},
  184110\relax
\mciteBstWouldAddEndPuncttrue
\mciteSetBstMidEndSepPunct{\mcitedefaultmidpunct}
{\mcitedefaultendpunct}{\mcitedefaultseppunct}\relax
\EndOfBibitem
\bibitem[Alkauskas \latin{et~al.}(2014)Alkauskas, Buckley, Awschalom, and
  Van~de Walle]{Alkauskas_2014}
Alkauskas,~A.; Buckley,~B.~B.; Awschalom,~D.~D.; Van~de Walle,~C.~G.
  First-principles theory of the luminescence lineshape for the triplet
  transition in diamond NV centres. \emph{New Journal of Physics}
  \textbf{2014}, \emph{16}, 073026\relax
\mciteBstWouldAddEndPuncttrue
\mciteSetBstMidEndSepPunct{\mcitedefaultmidpunct}
{\mcitedefaultendpunct}{\mcitedefaultseppunct}\relax
\EndOfBibitem
\bibitem[Wen and Dong(2024)Wen, and Dong]{Wen}
Wen,~F.; Dong,~G. Determining the molecular Huang-Rhys factor via
  scanning-tunneling- microscopy--induced luminescence. \emph{Phys. Rev. A}
  \textbf{2024}, \emph{109}, 042817\relax
\mciteBstWouldAddEndPuncttrue
\mciteSetBstMidEndSepPunct{\mcitedefaultmidpunct}
{\mcitedefaultendpunct}{\mcitedefaultseppunct}\relax
\EndOfBibitem
\bibitem[Grushka(1972)]{Grushka1970}
Grushka,~E. Characterization of exponentially modified Gaussian peaks in
  chromatography. \emph{Analytical Chemistry} \textbf{1972}, \emph{44},
  1733--1738\relax
\mciteBstWouldAddEndPuncttrue
\mciteSetBstMidEndSepPunct{\mcitedefaultmidpunct}
{\mcitedefaultendpunct}{\mcitedefaultseppunct}\relax
\EndOfBibitem
\bibitem[Parker~Jr \latin{et~al.}(1967)Parker~Jr, Feldman, and
  Ashkin]{parker1967raman}
Parker~Jr,~J.; Feldman,~D.; Ashkin,~M. Raman scattering by silicon and
  germanium. \emph{Physical Review} \textbf{1967}, \emph{155}, 712\relax
\mciteBstWouldAddEndPuncttrue
\mciteSetBstMidEndSepPunct{\mcitedefaultmidpunct}
{\mcitedefaultendpunct}{\mcitedefaultseppunct}\relax
\EndOfBibitem
\bibitem[Ciancico \latin{et~al.}(2025)Ciancico, Torre, Terrés, Moreno, Smit,
  Watanabe, Taniguchi, Orrit, Koppens, and
  Reserbat-Plantey]{SI-Ciancico-2025-ChargeDefects}
Ciancico,~C.; Torre,~I.; Terrés,~B.; Moreno,~A.; Smit,~R.; Watanabe,~K.;
  Taniguchi,~T.; Orrit,~M.; Koppens,~F.; Reserbat-Plantey,~A. Optical detection
  of charge defects near a graphene transistor using the Stark shift of
  fluorescent molecules. \emph{APL Quantum} \textbf{2025}, \emph{2},
  036101\relax
\mciteBstWouldAddEndPuncttrue
\mciteSetBstMidEndSepPunct{\mcitedefaultmidpunct}
{\mcitedefaultendpunct}{\mcitedefaultseppunct}\relax
\EndOfBibitem
\bibitem[Kulzer \latin{et~al.}(1997)Kulzer, Kummer, Matzke, Br{\"a}uchle, and
  Basch{\'e}]{kulzer1997single}
Kulzer,~F.; Kummer,~S.; Matzke,~R.; Br{\"a}uchle,~C.; Basch{\'e},~T.
  Single-molecule optical switching of terrylene in p-terphenyl. \emph{Nature}
  \textbf{1997}, \emph{387}, 688--691\relax
\mciteBstWouldAddEndPuncttrue
\mciteSetBstMidEndSepPunct{\mcitedefaultmidpunct}
{\mcitedefaultendpunct}{\mcitedefaultseppunct}\relax
\EndOfBibitem
\bibitem[Bordat and Brown(2000)Bordat, and Brown]{bordat2000elucidation}
Bordat,~P.; Brown,~R. Elucidation of optical switching of single guest
  molecules in terrylene/p-terphenyl mixed crystals. \emph{Chem. Phys. Lett}
  \textbf{2000}, \emph{331}, 439--445\relax
\mciteBstWouldAddEndPuncttrue
\mciteSetBstMidEndSepPunct{\mcitedefaultmidpunct}
{\mcitedefaultendpunct}{\mcitedefaultseppunct}\relax
\EndOfBibitem
\bibitem[Xu \latin{et~al.}(2018)Xu, Elbadawi, Tran, Kianinia, Li, Liu, Hoffman,
  Nguyen, Kim, Edgar, Wu, Song, Ali, Ford, Toth, and
  Aharonovich]{xu2018singlephoton}
Xu,~Z.-Q. \latin{et~al.}  Single photon emission from plasma treated 2D
  hexagonal boron nitride. \emph{Nanoscale} \textbf{2018}, \emph{10},
  7957--7965\relax
\mciteBstWouldAddEndPuncttrue
\mciteSetBstMidEndSepPunct{\mcitedefaultmidpunct}
{\mcitedefaultendpunct}{\mcitedefaultseppunct}\relax
\EndOfBibitem
\bibitem[Zeng \latin{et~al.}(2024)Zeng, Zhang, Meng, Chen, Jiang, Shi, Huang,
  Yin, Wu, and Zhang]{zeng2024single}
Zeng,~L.; Zhang,~S.; Meng,~J.; Chen,~J.; Jiang,~J.; Shi,~Y.; Huang,~J.;
  Yin,~Z.; Wu,~J.; Zhang,~X. Single-Photon Emission from Point Defects in
  Hexagonal Boron Nitride Induced by Plasma Treatment. \emph{ACS Applied
  Materials \& Interfaces} \textbf{2024}, \emph{16}, 24899--24907\relax
\mciteBstWouldAddEndPuncttrue
\mciteSetBstMidEndSepPunct{\mcitedefaultmidpunct}
{\mcitedefaultendpunct}{\mcitedefaultseppunct}\relax
\EndOfBibitem
\bibitem[Lewis(1995)]{lewis1995fast}
Lewis,~J.~P. Fast Template Matching. Vision Interface. Quebec City, Canada,
  1995; pp 120--123\relax
\mciteBstWouldAddEndPuncttrue
\mciteSetBstMidEndSepPunct{\mcitedefaultmidpunct}
{\mcitedefaultendpunct}{\mcitedefaultseppunct}\relax
\EndOfBibitem
\bibitem[Avants \latin{et~al.}(2008)Avants, Epstein, Grossman, and
  Gee]{avants2008symmetric}
Avants,~B.~B.; Epstein,~C.~L.; Grossman,~M.; Gee,~J.~C. Symmetric diffeomorphic
  image registration with cross-correlation: evaluating automated labeling of
  elderly and neurodegenerative brain. \emph{Medical Image Analysis}
  \textbf{2008}, \emph{12}, 26--41\relax
\mciteBstWouldAddEndPuncttrue
\mciteSetBstMidEndSepPunct{\mcitedefaultmidpunct}
{\mcitedefaultendpunct}{\mcitedefaultseppunct}\relax
\EndOfBibitem
\bibitem[Friedman \latin{et~al.}(1977)Friedman, Bentley, and
  Finkel]{friedman1977algorithm}
Friedman,~J.~H.; Bentley,~J.~L.; Finkel,~R.~A. An algorithm for finding best
  matches in logarithmic expected time. \emph{ACM Transactions on Mathematical
  Software (TOMS)} \textbf{1977}, \emph{3}, 209--226\relax
\mciteBstWouldAddEndPuncttrue
\mciteSetBstMidEndSepPunct{\mcitedefaultmidpunct}
{\mcitedefaultendpunct}{\mcitedefaultseppunct}\relax
\EndOfBibitem
\bibitem[Adhikari \latin{et~al.}(2024)Adhikari, Smit, and Orrit]{Orrit_2024}
Adhikari,~S.; Smit,~R.; Orrit,~M. Future Paths in Cryogenic Single-Molecule
  Fluorescence Spectroscopy. \emph{The Journal of Physical Chemistry C}
  \textbf{2024}, \emph{128}, 3--18\relax
\mciteBstWouldAddEndPuncttrue
\mciteSetBstMidEndSepPunct{\mcitedefaultmidpunct}
{\mcitedefaultendpunct}{\mcitedefaultseppunct}\relax
\EndOfBibitem
\end{mcitethebibliography}

}

\begin{document}

\begin{abstract}
Single polycyclic aromatic hydrocarbon molecules embedded in {organic} matrices have proven to be an excellent family of narrow-linewidth quantum emitters. Extending this host-guest setting to van der Waals materials offers the opportunity to combine the preeminent properties of molecular emitters with the access to the versatility of two-dimensional hetero-structures and devices. In this work, we incorporate perylene molecules into multi-layered hexagonal boron nitride stacks and observe gigahertz-narrow zero-phonon-line transitions at cryogenic temperatures. We unambiguously verify the origins of photon emission through vibronic spectra analysis. By combining hyperspectral localization measurements with quantum chemistry calculations, we examine the insertion mechanisms of perylene molecules in the hexagonal boron nitride stacks, and conclude that {pristine hBN layers tend to expel molecules from the sandwich, while extended morphological defects, hydroxyl groups and unpassivated boron and nitrogen atoms assist to stabilize molecular bindings to hBN}. Our work provides valuable insight for future work to deterministically integrate narrow-linewidth molecular emitters into van der Waals devices. 
\end{abstract}

\section{Introduction}
As a wide band-gap two-dimensional material, hexagonal boron nitride (hBN) has recently gathered tremendous interest as a host material of single-photon emitters\,\cite{Aharonovich-2022-review,Im-2025-Review}. hBN can be easily exfoliated down to few layers, transferred onto a wide range of substrates, and incorporated into heterostructures and hybrid devices\,\cite{Geim-2013-vdW}. A plethora of single-photon emitters in hBN has been discovered in the past decade\,\cite{Cholsuk-2024-Datebase}. Some of them exhibit remarkable properties such as photon indistinguishability\,\cite{Fournier-2023-TwoPhoton}, room-temperature operation\,\cite{Chatterjee-2025-RoomTemp}, and optically-based spin initialization and readout\,\cite{Gottscholl-2020-Initialization,Stern-2022-ODMR,Gao-2025-SingleNuclearSpin}. Yet, apart from a few species which could be reproducibly created or traced to identifiable atomic-scale structure\,\cite{Gottscholl-2020-Initialization,Tang-2025-StructuredDefect, Hua-2025-Deterministic}, unambiguous assignment of the microscopic origin of most emitters remains an outstanding challenge\,\cite{Carbone-2025-Microscopic}.

Concurrently, polycyclic aromatic hydrocarbon (PAH) molecules have established themselves as a class of well-understood and characterized single-photon emitters\,\cite{Toninelli-2021-single, Gurlek-2025-Small}. They exhibit tunable, nearly Fourier-limited single-photon emission when embedded into organic matrices and cooled to liquid helium temperatures and are integrable into a wide range of photonic devices\,\cite{Gurlek-2025-Small}. The adaptability of PAH emitters into hBN has sparked recent attention\,\cite{Wang-2025-IdealSubstrates}. Dibenzoterrylene-doped anthracene nanocrystals have been integrated into hetero-structures, where narrow-linewidth and electrically tunable single-photon emissions have been demonstrated\,\cite{Schaedler-2019-Electrical}. The same architecture has been extended to enable the optical detection of trapped charges near a graphene transistor\,\cite{Ciancico-2025-ChargeDefects}. Further recent studies have reported improvement of photostability of terrylene adsorbed on the surface of hBN\,\cite{Han-2021-Photostable}, and of {TIPS}-pentacene emitters placed on fused silica ($\mathrm{SiO_2}$) substrate and encapsulated by hBN\,\cite{Farooqui-2025-TIPS}, both at room temperature. At liquid helium temperature, narrow optical transitions down to 400\,MHz have been measured on {single} terrylene {molecules} adsorbed on the hBN surface\,\cite{Smit-2023-Sharp, deHaas-2025-Charge}. On a different front, a certain family of hBN emitters {with} the emission energy {at around} 2\,eV has been systematically verified to originate from organic molecules introduced by solvents and trapped between hBN and the $\mathrm{SiO_2}$ surface\,\cite{Neumann-2023-Organic}.

These recent developments motivate the research question whether PAH emitters can be directly inserted into hBN stacks without the need for nanoscale organic hosts and whether lifetime-limited molecular transitions can be detected. Answering these questions would open up possibilities for engineering hBN-based single-photon sources that are chemically tunable across a broad range of wavelengths, and proximity-integrated molecular probes for studying local charge and strain dynamics. In seeking answers to these questions, we report a first, combined experimental and theoretical investigation on {controlled incorporation of PAH molecules into hBN stacks}.

\section{Integrating PAH molecules into hBN}
\begin{figure}
\includegraphics[scale=1]{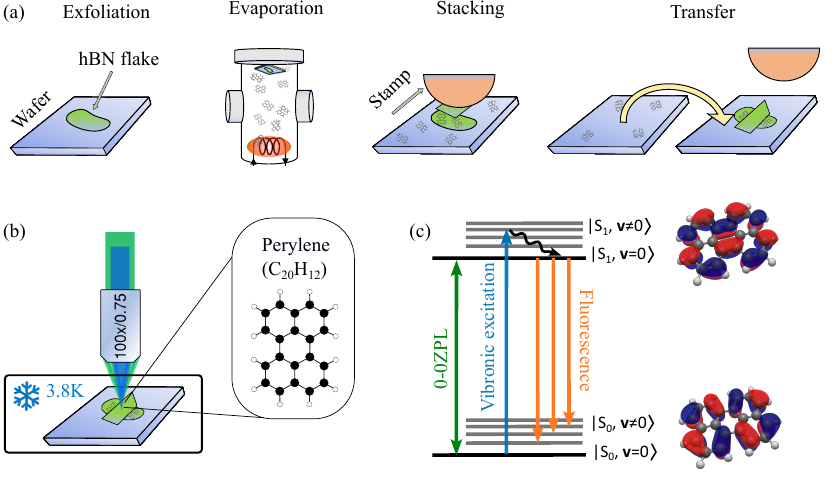}
\caption{(a) Sample fabrication routine. An hBN flake is first prepared on a silicon wafer capped with a $\text{Si}\text{O}_{2}$ layer. The substrate is then transferred into the evaporation chamber. After deposition, a second hBN flake is picked up by a polymer dome and stacked on the deposited hBN flake. The stack is picked up and transferred to a clean silicon substrate. (b) The sample is transferred to a closed-cycle cryostat, cooled to 3.8\,Kelvin and studied under an optical microscope. Inset shows the molecular structure of perylene. (c) A simplified energy level diagram of perylene. The two images on the right side display the $\pi$-orbitals involved in the $|S_0\rangle$ to $|S_1\rangle$ transition.}
\label{fig:sample}
\end{figure}

Specifically, we chose perylene ($\text{C}_{20}\text{H}_{12}$), a small PAH molecule that emits in the blue, offset from the wavelength range of most common hBN emitters. In addition, perylene has been studied in various single-molecule experiments\,\cite{Walla-1998-Perylene,Verhart-2016,Fang-2021,Smit-2022-Reverse} and features a comparatively {non-complex} vibrational energy structure. We introduce perylene molecules to hBN through thermal evaporation. As shown in Fig.~\ref{fig:sample}, commercial hBN flakes purchased from 2D Semiconductors are first prepared on a silicon substrate (terminated with a $\mathrm{SiO_2}$ layer) through exfoliation. Perylene molecules are dissolved in chloroform at a mass ratio of 1:3100. A nickel-chromium alloy filament is immersed into the solution, dried, and then fixed on the electric feed-through of a vacuum flange, which is then attached to an inert gas chamber. A current passing through the filament produces ohmic heating and deposits perylene molecules onto the hBN flake ({see SI section 1 for details about the evaporation process and sources of materials used}). After evaporation, a second hBN flake, exfoliated on a clean substrate is picked up by a stamp made of polydimethylsiloxane (PDMS) with a polycarbonate (PC) sacrificial layer and stacked onto the evaporated flake while the substrate is heated up to \SI{130}{\celsius}. For samples with high evaporation doses, the stack containing embedded molecules is picked up by the same stamp and transferred to a fresh substrate in order to isolate signals from molecules on the bare $\mathrm{SiO_2}$ surface. The stamp is detached by heating up the substrate to \SI{170}{\celsius}, leaving the stack and the PC layer on the substrate. After removing the PC layer with chloroform and cleaning with acetone, ethanol, and isopropanol, the substrate holding the stack is transferred into a closed-cycle helium cryostat and cooled to below 4\,Kelvin for optical measurements. {An additional baking step at up to \SI{750}{\celsius} is applied to one sample to enable a controlled study of the effect of thermal annealing. A list of all samples and their processing steps is provided in SI section 1}.

The photophysics of perylene molecules embedded in a solid is illustrated by the Jablonski diagram Fig.\,\ref{fig:sample}(c). The electronic ground state $|S_0\rangle$ and the lowest singlet excited state $|S_1\rangle$ facilitate a strong optical transition, corresponding to the promotion of a $\pi$-electron to an excited orbital. Each of the two electronic states splits into a manifold $|S_{0,1},\bm{\nu}\rangle$, where $|\bm{\nu}=\prod_i\nu^{(i)}\rangle$ denotes the vector space supported by the vibrational modes of the molecule, with $i$ the mode index and $\nu^{(i)}=0,1,2...$ the quantum number of occupation. The molecule can be optically excited from the vibronic ground state $|S_0,\bm{\nu}=\bm{0}\rangle$ to a vibrational level of the excited state (blue arrow), from where it rapidly relaxes to $|S_1,\bm{\nu}=\bm{0}\rangle$ (black curly arrow). The transition between $|S_0,\bm{\nu}=\bm{0}\rangle$ and $|S_1,\bm{\nu}=\bm{0}\rangle$ is called the 0-0 zero-phonon-line transition (0-0 ZPL, green arrow), and for simplicity, we refer to it as ZPL. Emissions to all other sublevels in the ground state manifold are allowed (orange arrows) with their relative strength described by the Franck-Condon factors. Measuring energy spacings among fluorescence emission lines thus allows reconstruction of the vibrational energy landscape of the ground-state manifold and enables unambiguous identification of the origin of optical emission.

\begin{figure}[!htbp]
\includegraphics[scale=0.38]{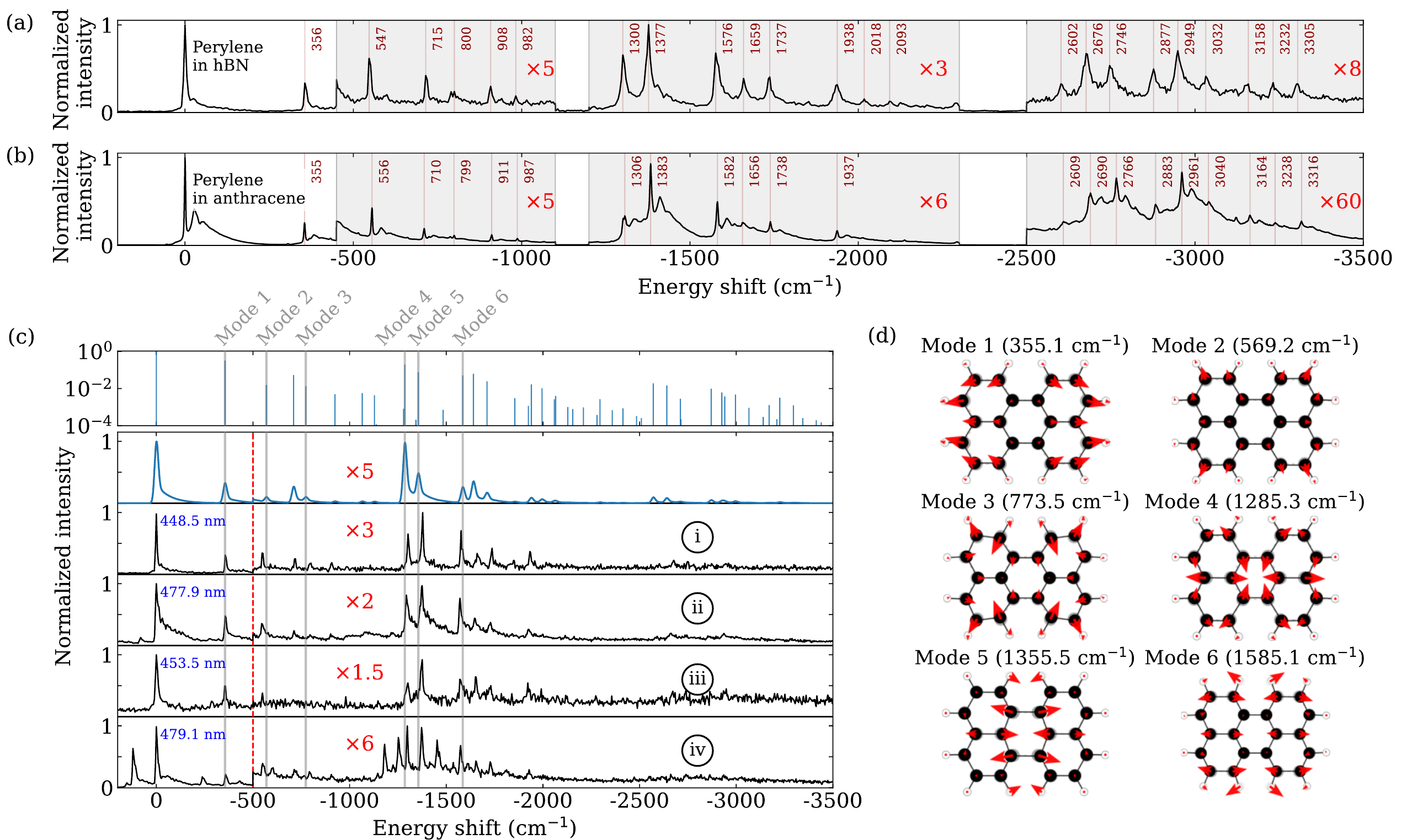}
\caption{{Fluorescence spectrum measured from a focal spot on the hBN stack (a) compared to the emission spectrum of perylene in anthracene (b). The horizontal axis denotes the relative energy shift to the ZPL, expressed in inverse centimeters. The vibronic bands are scaled differently in intensity for visual effect. {(c)} Comparison to theory.} The blue vertical bars in the top panel present the vibrational modes of perylene in an hBN slab. The height of each bar represents the Franck-Condon factors derived from \textit{ab initio} calculations. Only six modes and their overtones are plotted. The blue curve shows a modeled emission spectrum, which considers a broadening by the spectrometer and the phonon sideband. A universal energy scaling of 1.1 is performed to the theory data. The four spectra (i) to (iv) are measured at various locations on different samples. Spectrum {(iv)} features two sets of emission peaks, signifying that there are at least two molecules located in the optical focus. {(d)} Cartoons of the six vibrational modes and their calculated energy. The length of arrows illustrates the amplitude of the in-plane component of the acceleration experienced by each atom.}
\label{fig:spectra}
\end{figure}

We start with vibronic excitation of the sample {with the laser frequency} blue detuned with respect to the ZPL wavelengths of perylene measured in organic matrices\,\cite{Walla-1998-Perylene,Verhart-2016}. The stacked area is imaged on a scientific CMOS camera, from which isolated fluorescent spots can be identified. Fluorescence light from each spot is spatially filtered in a confocal optical setup, coupled to a fiber and guided to a grating spectrometer. {See SI section 1 for the camera images and section 2 for details of the optical setup.} The signal shown in Fig.\,\ref{fig:spectra}(a) displays a representative emission {spectrum. It features} a series of dispersed sharp peaks, each accompanied by a sideband on the lower energy side. The sharp peaks resemble molecular emission features from $|S_1,\bm{\nu}=\bm{0}\rangle$ to different $|S_0,\bm{\nu}\rangle$ sub-levels, while the sideband associated with each peak is known as the phonon sideband induced by coupling to phonons in the host material\,\cite{Gurlek-2025-Small}. 
To validate that the emission is from perylene, {we compare this spectrum with a reference measured from perylene molecules doped into the organic host anthracene, as shown in Fig.\,\ref{fig:spectra}(b). All identifiable vibronic peaks are in excellent agreement. The relative peak intensities and the shape of the phonon sidebands differ between the two cases, signifying different equilibrium geometries of perylene and different phonon densities of states in the two hosts. To further understand the microscopic origin of each peak,} we calculate with the ORCA code\cite{neese2025} the excited-state relaxation of perylene using time-dependent density-functional theory (B3LYP D4 def2-TZVPD)\cite{lee1988development,becke1993density,caldeweyher2019generally,casida1996time} and estimate the vibrational modes of the hBN slabs using the universal machine-learning potential MACE-MP\,\cite{batatia2023foundation}. The vibrational mode frequencies and the associated Franck-Condon factors are extracted from this combination and displayed by the blue vertical bars in the top panel of {Fig.\,\ref{fig:spectra}(c)}. To obtain a clearer visual comparison, we convolve the peaks with the spectral response function of the spectrometer and an exponential function modeling the phonon sidebands (details about the model are given in section~4 of the SI). The resulting spectrum after a universal energy scaling by a factor of 1.109 is displayed by the blue curve. {Shown below the theory curve are four additional perylene spectra measured from different samples.} We find that almost all peaks can be explained by six modes {and their overtones} with up to three excitation quanta in each mode. The sketches in {Fig.\,\ref{fig:spectra}(d)} illustrate these six vibrational modes. It is to note that spectrum {(iv)} consists of doublets of each vibronic transition and thus likely originates from two perylene molecules in the same confocal spot.

Next, we extract the energies of these six modes from the measured spectra and compare them with the corresponding values obtained from {our measurements of perylene in anthracene, as well as with literature values reported for perylene in biphenyl}\,\cite{Walla-1998-Perylene}. The results are displayed in Tab.\,\ref{tab:perylene_modes}. {The extracted mode energies agree well with the reference values of perylene in anthracene and biphenyl, with a root-mean-squared deviation of approximately $7.9\,\mathrm{cm^{-1}}$, corresponding to an averaged deviation below 1\%. At the same time, modes 2 to 6 are systematically shifted to lower energies by about 7 to 9\,$\mathrm{cm^{-1}}$, suggesting that the hBN environment induces a slight softening of these modes.} These comparisons lead us to conclude that the observed emission features indeed originate from perylene molecules.

\begin{table}[ht]
\centering
\small
\setlength{\tabcolsep}{4pt}
\begin{tabular}{ccccccccc}
\hline
Mode &
Biphenyl &
Anthracene &
M-i &
M-ii &
M-iii &
M-iv.1 &
M-iv.2 &
DFT Calculation \\
Number &
(cm$^{-1}$) &
(cm$^{-1}$) &
(cm$^{-1}$) &
(cm$^{-1}$) &
(cm$^{-1}$) &
(cm$^{-1}$) &
(cm$^{-1}$) &
(cm$^{-1}$) \\
\hline
1 & 358 & 355 & 356.4 & 357.4 & 355.9& 358.5 & 360 & 355.1 \\
2 & 557 & 556 & 549 & 546.5 & 549 & 550 & 553.5 & 569.2 \\
3 & 808 & 799 & 796.5 & 795.5 & -- & 791.4 & -- & 773.5 \\
4 & 1311 & 1306 & 1302.1 & 1293.5 & 1301.2 & 1298 & 1302 & 1285.3 \\
5 & 1384 & 1383 & 1378 & 1374.2 & 1373.7 & 1374 & 1375 & 1355.5 \\
6 & 1584 & 1582 & 1577 & 1570.8 & 1572.6 & 1574 & 1574.5 & 1585.1 \\
\hline
\end{tabular}
\caption{Vibrational mode energies of perylene in anthracene and biphenyl compared to the extracted values from the spectra in Fig.\,\ref{fig:spectra}(c) and the calculated values with MACE-MP potential after a global scaling. M-i to M-iv.2 denote the five sets of molecular emission features extracted from the four spectra.}
\label{tab:perylene_modes}
\end{table}

\section{{Spectral stability}}
\begin{figure}
\includegraphics[scale=0.7, trim=1.5cm 0cm 0cm 0cm,clip]{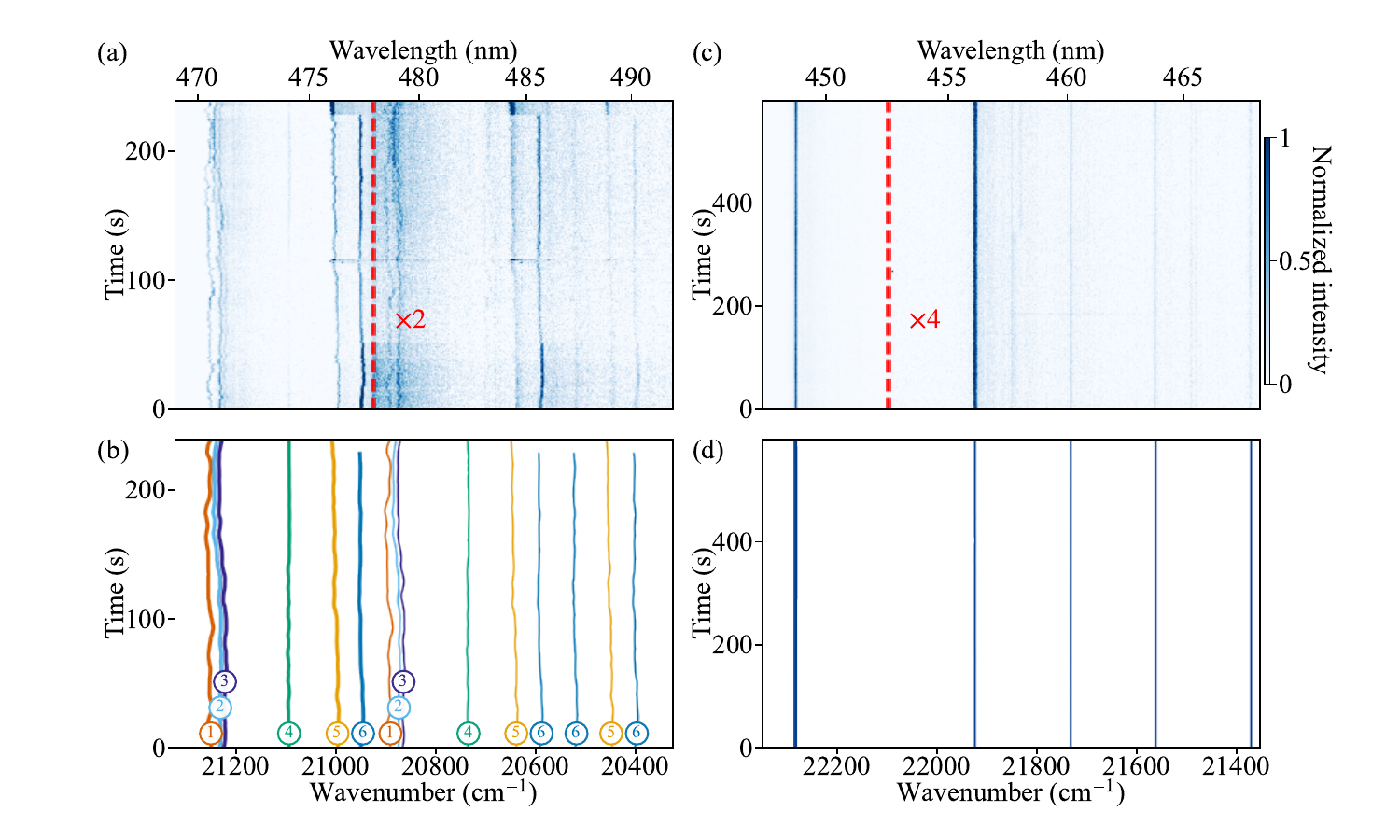}
\caption{(a) Emission spectrum measured on a laser focal spot {on a non-baked sample} over 200 seconds. (b) Extracted frequency traces of 6 molecules. Each molecule is represented by one color. The opaque lines denote the ZPL transition, and the semi-transparent lines are the vibronic transitions. (c) {An emission spectrum measured on a sample after baking. (d) Extracted time traces of the ZPL and vibronic lines for the data in (c).}}
\label{fig:diff}
\end{figure}

Next, we study the spectral stability of the molecules, which can shed light on the stability of their local nanoscopic environment. Figure\,\ref{fig:diff}(a) displays the evolution of the emission spectrum obtained from a diffraction-limited fluorescence spot over a continuous measurement time of 220 seconds. {This sample has not gone through thermal annealing}. A careful analysis of the spectra (details in SI section 5) suggests that six separate perylene emission features are present. Figure\,\ref{fig:diff}(b) shows the trajectories of the ZPL transition (opaque lines) of the molecules and their associated vibronic lines (semi-transparent lines). The three frequency traces labelled 1-3 on the left side of the plot all exhibit strong spectral diffusion. {The vibronic emission lines shift synchronously with their respective ZPLs. This suggests that the spectral shifts are due to electronic coupling to their environment, instead of changes involving mechanical distortions, such as hopping between binding sites. The latter would lead to pronounced modifications to the vibrational mode energies and Franck-Condon factors. Further observations show that molecules 2 and 3 exhibit very similar diffusion patterns, which differ from that of molecule 1, indicating that molecules 2 and 3 experience similar local environments that are different from the environment of molecule 1.} Molecule 4 has remained mostly stable. Molecules 5 and 6 show interesting correlations in their emission intensity. Towards the end of the measurement sequence, molecule 6 suffered from photobleaching. At the same time, an increase in intensity appears in the ZPL and vibronic lines of molecule 5. {This observation can be explained by F{\"o}rster resonance energy transfer (FRET). Molecule 5, having higher energy, functions as a donor and has certain probability to transfer its excitation to the energetically lower-lying molecule 6, the acceptor. Photobleaching of organic fluorophores is associated with the breaking of the $\pi$-delocalization via a chemical reaction, which leads to a significant increase in the absorption energy gap\,\cite{Mhanna-2025}. When the acceptor molecule is bleached, its excited-state energy shifts above that of the donor. The resonance condition of FRET is no longer fulfilled, and the fluorescence quantum yield of molecule 5 increases, leading to its increase in brightness. Since the efficiency of FRET is inversely proportional to the 6th power of the intermolecular distance\,\cite{Hofkens-2003-FRET}, our observation that molecule 5 becomes approximately a factor of 2 brighter after molecule 6 has undergone photobleaching implies a donor-acceptor separation close to the Förster radius, which should amount to several nanometers.}

{The observation of photobleaching suggests the presence of reactive species in the local intermediate of the molecules. During sample preparation, water and oxygen molecules can be trapped within the sandwich, or adsorbed at defect sites, and can thus act as reactants leading to oxidation of the acceptor molecule. In addition, the rotation or diffusion of these small molecules may contribute to spectral diffusion of the perylene. To test this hypothesis, we perform a control experiment by baking one of the hBN stacks at \SI{750}{\celsius} in air for a duration of 4 hours. At this temperature, we expect that small molecules like water or oxygen can still diffuse through and escape the stack. Shown in Fig.\,\,\ref{fig:diff}(c) is a perylene emission spectrum measured on the baked sample, and Fig.\,\,\ref{fig:diff}(d) illustrates the respective spectral traces of the ZPL and vibronic lines. Here, we observe no significant intensity change and spectral drift over the duration of 10 minutes. This observation supports the conclusion that removing small adsorbate molecules reduces environmental fluctuations and thereby creates a more stable local environment.}

\begin{figure}
\includegraphics[scale=0.8]{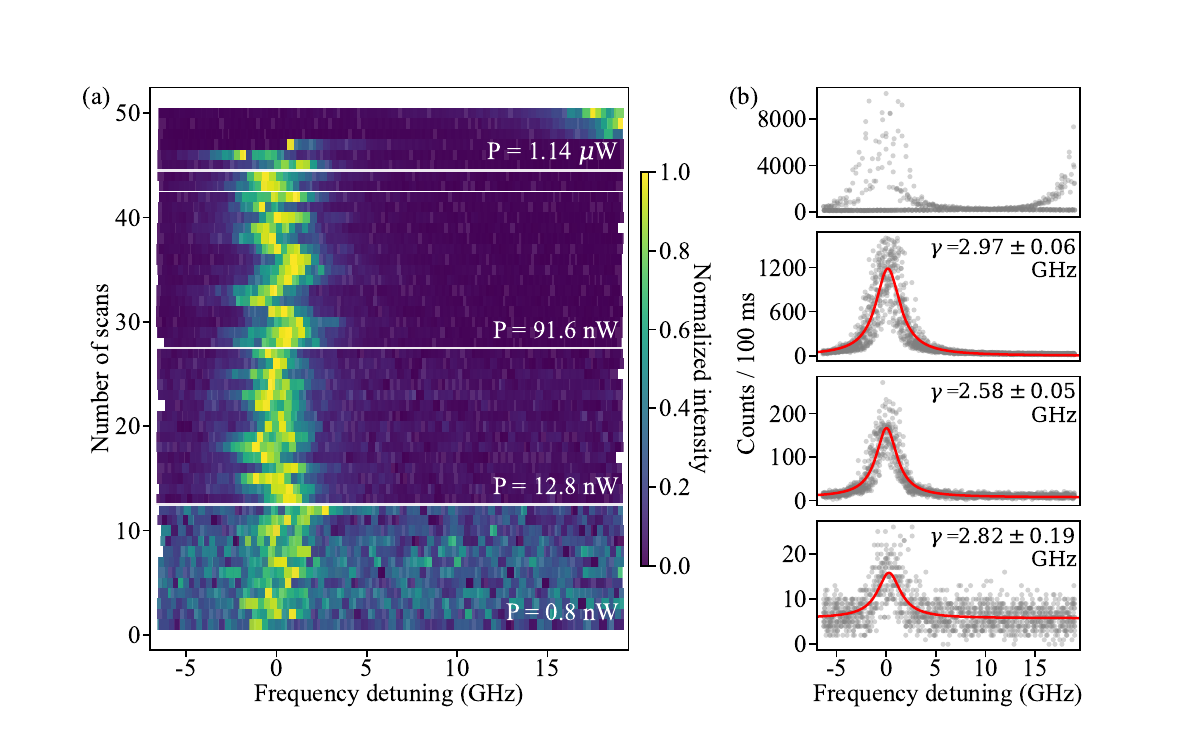}
\caption{Fluorescence excitation spectra of the ZPL transition of a single perylene molecule. (a) The ZPL excitation traces measured with gradually increased laser intensities. The duration of a single frequency sweep was 12 seconds. The color scale is normalized in each scan. Spectral jumps occur at the highest excitation power. (b) Excitation spectra at each laser power. The gray dots show the accumulated data from (a) without post-aligning the center resonance frequency of each individual scan. The red lines denote fits to a Lorentzian function. The extracted full-width at half-maximum linewidth $\gamma$ is given in the legend.}
\label{fig:zpl}
\end{figure}

Next, we perform resonant excitation to probe the linewidth of the ZPL transition. To do so, the excitation laser is tuned to 476\,nm. The fluorescence signal from the $|S_1,\bm{\nu}=\bm{0}\rangle$ to different $|S_0,\bm{\nu\neq 0}\rangle$ transitions is separated from the excitation light with a set of long-pass interference filters and measured by an avalanche photodetector, while the laser frequency is scanned across the ZPL transition of a molecule. Figure\,\ref{fig:zpl}(a) shows the measured fluorescence excitation spectra of a molecule. In this set of measurements, the excitation power is increased stepwise from {0.8\,\textrm{nW} to 1.14\,\textrm{$\mu$W}} (measured before the cryostat window). Figure\,\ref{fig:zpl}(b) presents the extracted excitation spectra at different powers. A linewidth of around 2.8\,GHz is observed without noticeable power broadening. As the laser intensity increases to 1.14\,\textrm{$\mu$W}, fluorescence counts of up to 80,000 per second can be detected. However, a {spectral} jump occurs at this laser power, which signifies a photo-induced change in the local environment of the molecule. The abrupt on-off change in the fluorescence intensity evidences that the feature originates from a single molecule. The observed linewidth is considerably broader than single perylene molecules measured in biphenyl (140\,MHz, Ref.\cite{Walla-1998-Perylene}) and \textit{n}-nonane (28\,MHz, Ref.\cite{Pirotta-1996}), implying a stronger pure dephasing experienced by the molecule in hBN. The linewidth dominated by pure dephasing also explains the absence of noticeable power broadening in the linewidths measured in Fig.\,\ref{fig:zpl}(b).

\section{Hyperspectral analysis and \textit{ab initio} calculations}
\begin{figure}
\includegraphics[scale=0.42]{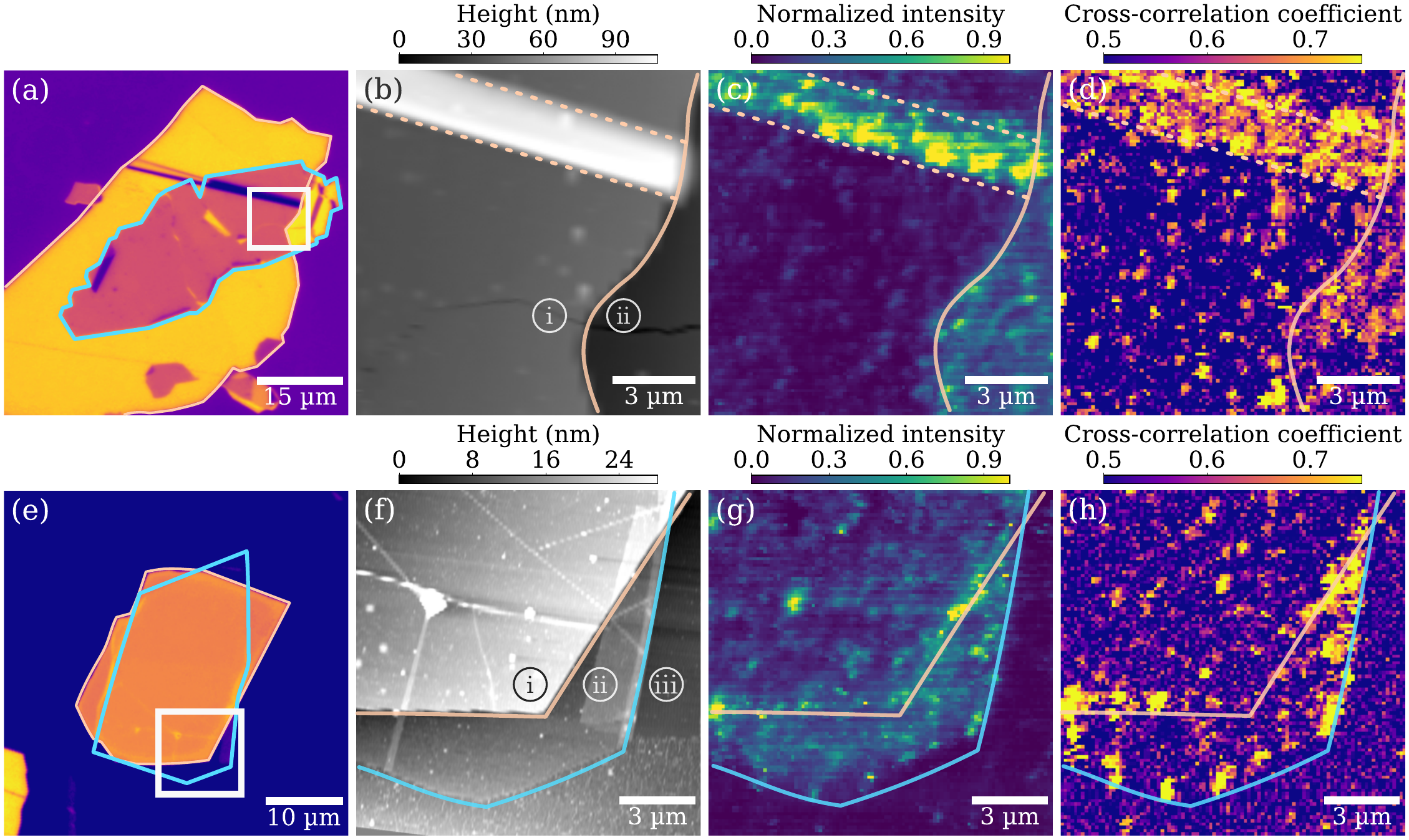}
\caption{Spatial distribution of perylene emissions analyzed through hyperspectral measurements on two samples. (a), (e) Optical microscope images of the two samples. Orange and blue contours illustrate the boundaries of the lower and upper hBN flakes, respectively. (b), (f) Atomic force microscope measurements. The top and bottom flakes in (b) have a thickness of 21\,nm and 31\,nm, respectively. The sample in (f) has a 22-nm-thick bottom flake and a 2-nm-thick top flake. (c), (g) Integrated fluorescence intensity distributions. (d), (h) Normalized correlation function with respect to the emission spectrum of perylene.}
\label{fig:hyper}
\end{figure}

An important open question is about the insertion geometries of perylene molecules in the hBN stack. To gain experimental insight, we perform spatially-resolved emission spectrum measurements on the sample. To do so, the laser focus is raster scanned across the sample and an emission spectrum is acquired at each position. The results are analyzed in correlation with topographic measurements using an atomic force microscope (AFM). Figures\,\ref{fig:hyper}(a) and (e) show the white-light microscope images of two samples, where the boundaries of the bottom and top hBN flakes are marked with orange and blue lines, respectively. The area in the white square is studied. The results from the AFM measurements are shown in Figs.\,\ref{fig:hyper}(b) and (f), respectively. The bright stripes in both images are bulges formed during exfoliation and stacking. The labels \textrm{(i)}, \textrm{(ii)}, and \textrm{(iii)} denote areas of the hBN-stack, $\mathrm{SiO_2}$ surface covered by the top flake, and bare $\mathrm{SiO_2}$ surface, respectively. Figures\,\ref{fig:hyper}(c) and (g) display the fluorescence intensity maps obtained by integrating the spectrum measured at each position. One can readily identify that areas stacked between flat hBN flakes show weaker fluorescence intensity compared to areas with bulges, flake edges, and substrate areas that are covered by the top hBN layer. To specifically trace perylene signals, we use the vibronic fingerprints in the range of 1200\,$\text{cm}^{-1}$ to 1800\,$\text{cm}^{-1}$ of perylene to perform an intensity-normalized cross-correlation analysis with the hyperspectral data set (see SI section 7 for details). The resulting correlation maps are displayed in Figs.\,\ref{fig:hyper}(d) and (h). In the areas stacked between flat hBN layers, individual scattered spots are identifiable. The areas with major morphological irregularities, i.e., flake edges, and bulges, appear to be much brighter than the stacked area. These locations are known to host high densities of defects, which possibly offer stronger bindings for perylene molecules. These results thus suggest that the binding of perylene is primarily defect-enhanced. The areas on the $\mathrm{SiO_2}$ surface and covered by the top hBN layer [area (ii)] also feature a strong intensity in the correlation map, which suggests that the $\mathrm{SiO_2}$ surface offers more favorable bindings for molecules compared to pristine hBN. It is to note that the stacks in (a) and (e) were both transferred to a fresh substrate before the optical measurements. The observation of perylene molecules in these area indicates that these molecules were picked up by the top flake from the initial substrate during the detaching process.

{Through hyperspectral analysis, we also extract the distribution of ZPL wavelengths (see SI section 7, Fig.\,S21). Across all samples prepared, we identify perylene ZPLs in the range of 440\,nm to 500\,nm. The distribution can be primarily categorized into a main peak between 445\,nm and 470\,nm, and a broader, more sparsely distributed group between 470\,nm and 500\,nm. For comparison, the ZPL wavelengths of perylene molecules embedded in organic matrices lie in the range of 444\,nm to 454\,nm\,\cite{Adhikari-2024}. The broader ZPL distribution observed in the perylene-hBN system therefore indicates substantially more heterogeneous insertion configurations of perylene. The overall red shift of the ZPLs with respect to the values observed in organic host matrices is consistent with our spatially resolved observation that perylene binding is primarily enhanced by defects. Coupling to defect sites in hBN may differentially stabilize the electronic ground and excited states of perylene, leading to a reduction of the optical transition energy. A red-shift of the ZPL has also been observed for terrylene molecules adsorbed on hBN surfaces\,\cite{Smit-2023-Sharp}.}

{We seek to gain a microscopic understanding of possible binding geometries} by numerically relaxing the combined perylene-hBN structure. We model the system using the ASE software package\,\cite{hjorth2017atomic} in combination with the foundation machine-learning interatomic potential MACE-MP (medium size)\,\cite{batatia2023foundation} and explicitly account for dispersive interactions. We create various possible defects and relax all cell parameters and nuclear positions for 50,000 steps using the FIRE optimization algorithm. The stability on insertion is benchmarked by calculating the formation energy (or binding energy of perylene) $E_\text{f} = E_\text{Mol+hBN} - (E_\text{Mol}+E_\text{hBN})$.
The results are presented in Tab.~\ref{tab:formation_energy} and all corresponding structures are visualized in section 3 of the SI.

\begin{figure}
\includegraphics[scale=0.6]{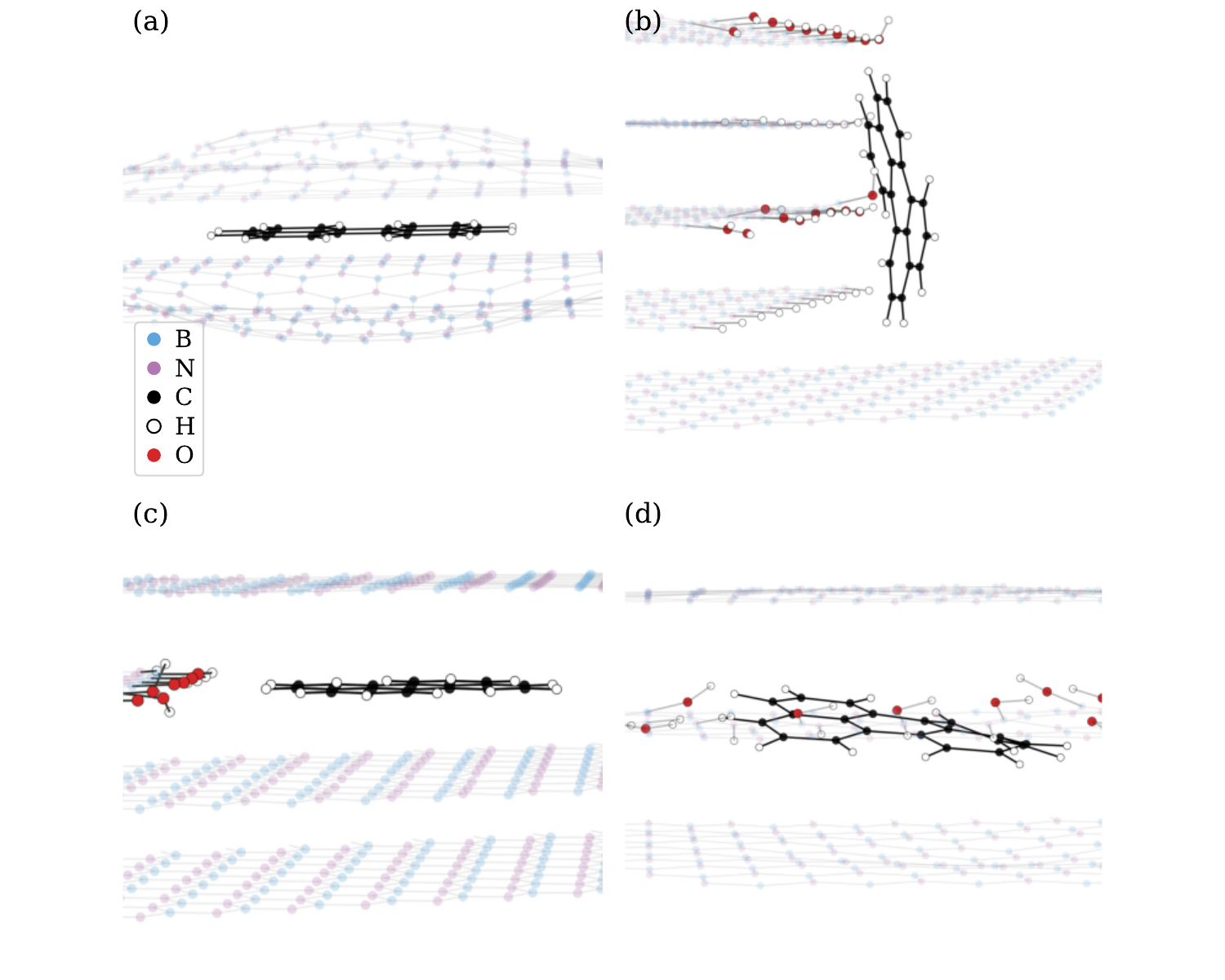}
\caption{Geometry of the relaxed molecule-hBN structure. (a) {E1}, perylene sandwiched in bi-layer hBN, (b) E10, multi-layer hBN edge, (c) E14, embedded step terrace, and (d) E16, extended vacancy. The structures in (b), (c) and (d) have -H terminated nitrogen and -OH terminated boron atoms.}
\label{fig:insertion}
\end{figure}

We found that sandwiching perylene between two pristine monolayers of hBN always leads to a positive formation energy in the order of 5.5\,eV [$\mathrm{E1}$ in Tab.\,\ref{tab:formation_energy} and Fig.\,\ref{fig:insertion}(a)], {suggesting that it is energetically expensive to sandwich perylene molecules between pristine layers.} Single point defects in the hBN lattice, including B- and N-vacancies, oxygen- or carbon substitutions ($\mathrm{E2-E5}$) lead to only marginal modifications to the formation energy. Extending the bilayer to multilayer hBNs increases the formation energy further ($\mathrm{E6-E7}$). {To estimate the energetic barrier for lateral motion of perylene within the stack, we perform a further set of simulations, where the lateral position of the perylene molecule is fixed along a path across a unit cell of hBN, while the hBN stack is allowed to relax. The results give an estimated migration barrier of $E_\mathrm{b}=54.6\,\mathrm{meV}$. This value is larger than the thermal energy at room temperature, which amounts to 25.9\,$\mathrm{meV}$ at 300\,$\mathrm{K}$, but not by orders of magnitude. The corresponding Arrhenius activation factor is $\exp(-E_\mathrm{b}/k_\mathrm{B}T) \approx 0.12$, suggesting that lateral motion is thermally accessible. During stacking and delamination, the sample is further heated, for which the activation factor increases to 0.24. The zipper-like closure of the interface can mechanically assist the displacement or expulsion of weakly bound molecules. These findings are} in accordance with previous studies indicating that the stacking process expels hydrocarbon molecules from hBN surfaces\,\cite{Haigh-2012-CrossSectionalImaging, Kretinin-2014-Electronic}.

We further explored the influence of extended defects, including multi-step edges [$\mathrm{E8-E10}$ and Fig.\,\ref{fig:insertion}(b)], embedded mono-step terraces [$\mathrm{E11-E14}$ and Fig.\,\ref{fig:insertion}(c)] and extended vacancies [$\mathrm{E15-E16}$ and Fig.\,\ref{fig:insertion}(d)]. Negative formation energies can be obtained in all these structures. {A further examination of the results shows that non-terminated boron atoms tend to covalently bind to carbon atoms in the molecule, and non-terminated nitrogen atoms stabilize the molecule through short-range interactions with the peripheral C--H groups of perylene.} This leads to rather large negative formation energies ($\mathrm{E8, E9, E11}$). {The formation energies for E8 and E9 could not be precisely captured by our numerical routine due to the random formation of} interlayer covalent bonds of boron and nitrogen atoms at the edge of the flake {(see SI section 3, Fig.\,S10)}. Physical bindings are found when the dangling boron and nitrogen atoms are terminated. Terminating boron atoms by hydroxyl groups ($\mathrm{E10, E14, E16}$) leads to considerably stronger binding than complete hydrogen terminations ($\mathrm{E12, E13}$), suggesting that the \textrm{-OH}$-\pi$ interaction\,\cite{Abelard-2016,Pohle-1982} potentially stabilizes the insertion of perylene. In our experiments, exfoliation and stacking are performed under ambient conditions. {Hydroxyl-terminated defects are expected to form more readily through reactions with water molecules than unpassivated radical sites or hydrogen-terminated defects. Direct covalent binding to the molecule, in particular B--C bond formation, would partially disrupt the $\pi$-delocalization of perylene. This would be expected to strongly modify both the optical transition energy and the vibrational spectrum, and is therefore unlikely to be captured by our hyperspectral analysis.}

The correlation map presented in Fig.\,\ref{fig:hyper}, in which perylene signals are primarily found at the edges and extended structural defects, is consistent with the calculation results. In addition, the areas sandwiched between hBN and $\mathrm{SiO_2}$ being rich of perylene signals is also reflected by the simulation results, as the $\mathrm{SiO_2}$ surface is known to be rich of hydroxyl terminations under exposure to humidity under ambient conditions\,\cite{Yang-2006}. The sparse distribution of perylene molecules in the stack area should be most likely due to extended vacancies and defects on the upper or lower hBN flake.

\begin{table}[htbp]
\centering
\footnotesize
\setlength{\tabcolsep}{4pt}
\renewcommand{\arraystretch}{1.15}
\begin{tabular}{c|l|l|c}
\hline
ID & Structure & Termination / defect & Formation energy (eV) \\
\hline
E1 & Bi-layer stack (1-Pr-1) & pristine & 5.548 \\
E2 & Bi-layer stack (1-Pr-1) & B vacancy & 5.151 \\
E3 & Bi-layer stack (1-Pr-1) & N vacancy & 5.764 \\
E4 & Bi-layer stack (1-Pr-1) & substitute N by O & 5.508 \\
E5 & Bi-layer stack (1-Pr-1) & substitute B by C & 5.528 \\
E6 & Multi-layer stack (4-Pr-4) & pristine & 8.215 \\
E7 & Multi-layer stack (4-Pr-4) & substitute N by O & 8.143 \\
E8 & Multi-step edge (4-Pr-4) & no termination & -- \\
E9 & Multi-step edge (4-Pr-4) & no termination & -- \\
E10 & Multi-step edge (4-Pr-4) & OH termination & $-1.114$ \\
E11 & Embedded step (3-Pr-1) & no termination & $-3.645$ \\
E12 & Embedded step (3-Pr-1) & H termination & $-0.770$ \\
E13 & Embedded step (3-Pr-2) & H termination & $-0.738$ \\
E14 & Embedded step (3-Pr-1) & OH termination & $-2.601$ \\
E15 & Extended vacancy (5-Pr-3) & no termination & 2.132 \\
E16 & Extended vacancy (5-Pr-3) & OH/H termination & $-3.940$ \\
\hline
\end{tabular}
\caption{Formation energies of perylene molecules inserted in pristine and defective hBN structures. The notation $m$-Pr-$n$ denotes a structure with perylene molecules binding to $m$-layers of bottom hBN and encapsulated with $n$ layers of top hBN. {Formation energies for $\mathrm{E8, E9}$ could not be precisely captured due to random formations of interlayer $\mathrm{B-N}$ bonds at flake edges.}}
\label{tab:formation_energy}
\end{table}

\section{Conclusion and discussion}

In this work, we performed a combined experimental and theoretical investigation of perylene molecules incorporated into hBN stacks. The choice of perylene, one of the smallest and well-characterized fluorophores in the PAH family allows us to unambiguously identify their vibronic emission fingerprints. The main conclusion of this work is twofold. First, cryogenic optical spectroscopy shows that perylene molecules can be unambiguously identified, and narrow zero-phonon-line optical transition down to 2.8\,GHz can be measured. Second, experimental results and theoretical calculations suggest that stable bindings of molecules in hBN should be primarily defect-enhanced. The results of hyperspectral analysis show that perylene insertion locations are strongly correlated to morphological irregularities in the structure, such as flake edges and wrinkles. {These findings are consistent with the self-cleaning mechanism proposed for stamp-based van der Waals assembly, in which weakly bound hydrocarbon adsorbates are expelled from pristine interfaces\,\cite{Haigh-2012-CrossSectionalImaging, Kretinin-2014-Electronic}. In this context, defects in hBN can act as local trapping sites that stabilize molecules against expulsion\cite{Lvova-2016-Theoretical}.} 

Our work can be viewed as a first step towards deterministic integration of molecular emitters into hBN. A line of future work is to be performed to make the system attractive for coherent quantum optical experiments. First, although narrow ZPL transition at the gigahertz level is observed, the holy grail of achieving a lifetime-limited linewidth, which is necessary for implementing coherent quantum protocols, still requires further investigation. To improve in this aspect, microscopic understanding of pure dephasing and spectral diffusion mechanisms is required. Here, insights may be gained from works on native hBN defects, where charge noise was found to be a main leading mechanism\,\cite{Akbari-2022-Modulation}. Future effort could incorporate electric gating structures to regulate the local electric field environment at the molecule. {Second, the density of isolated single molecules is much lower than in typical organic host matrices. The two-dimensional geometry provides fewer possible insertion sites than a three-dimensional crystal, and the stacking process likely expels most perylene molecules from pristine regions of the interface. Moreover, perylene is relatively volatile among PAHs and can exhibit high surface mobility, making systematic control of the molecular density challenging. In future experiments, larger molecules with lower surface mobility, such as PTCDA or TIPS-pentacene, could be used. Improved dose control may also be achieved by physical vapor deposition under vacuum\,\cite{Marquardt-2021-Homogeneous}.} In addition, controlled introduction of molecules may be achieved by controlling the location {and density} of extended defects, which can be achieved through e.g., surface self-assembly\,\cite{Fernez-2022-SelfAssembly}, or nano-indentation\,\cite{Luo-2025-Nanoindentation} prepared hBN surfaces. Finally, the good agreement between theory and experiment on perylene suggests that predictive theoretical schemes in combination with global optimization techniques hold potential to uncover the vast chemical space of molecular emitters for predictable integration into two-dimensional materials\,\cite{Ohman-2025-prediction}.

\begin{acknowledgement}

We thank Kathrin Schumacher for her assistance in the early stage of the experiment, and Moritz Sokolowski for suggesting the molecular deposition method. The work is supported by Germany’s Excellence Strategy - Cluster of Excellence Matter and Light for Quantum Computing (ML4Q) EXC 2004/2-390534769 and the European Union (ERC, MSpin, 101077866), and the TRA Matter at the University of Bonn as part of the Excellence Strategy of the federal and state governments.

\end{acknowledgement}

\CombinedMainBibliography

\StartCombinedSI

\section{Sample preparation}

\subsection{Evaporation of perylene molecules}\label{mol_evap}

\begin{figure}[h!]
    \centering
    \includegraphics[width=0.6\linewidth,trim=1cm 1.75cm 0.7cm 0.5cm,clip]{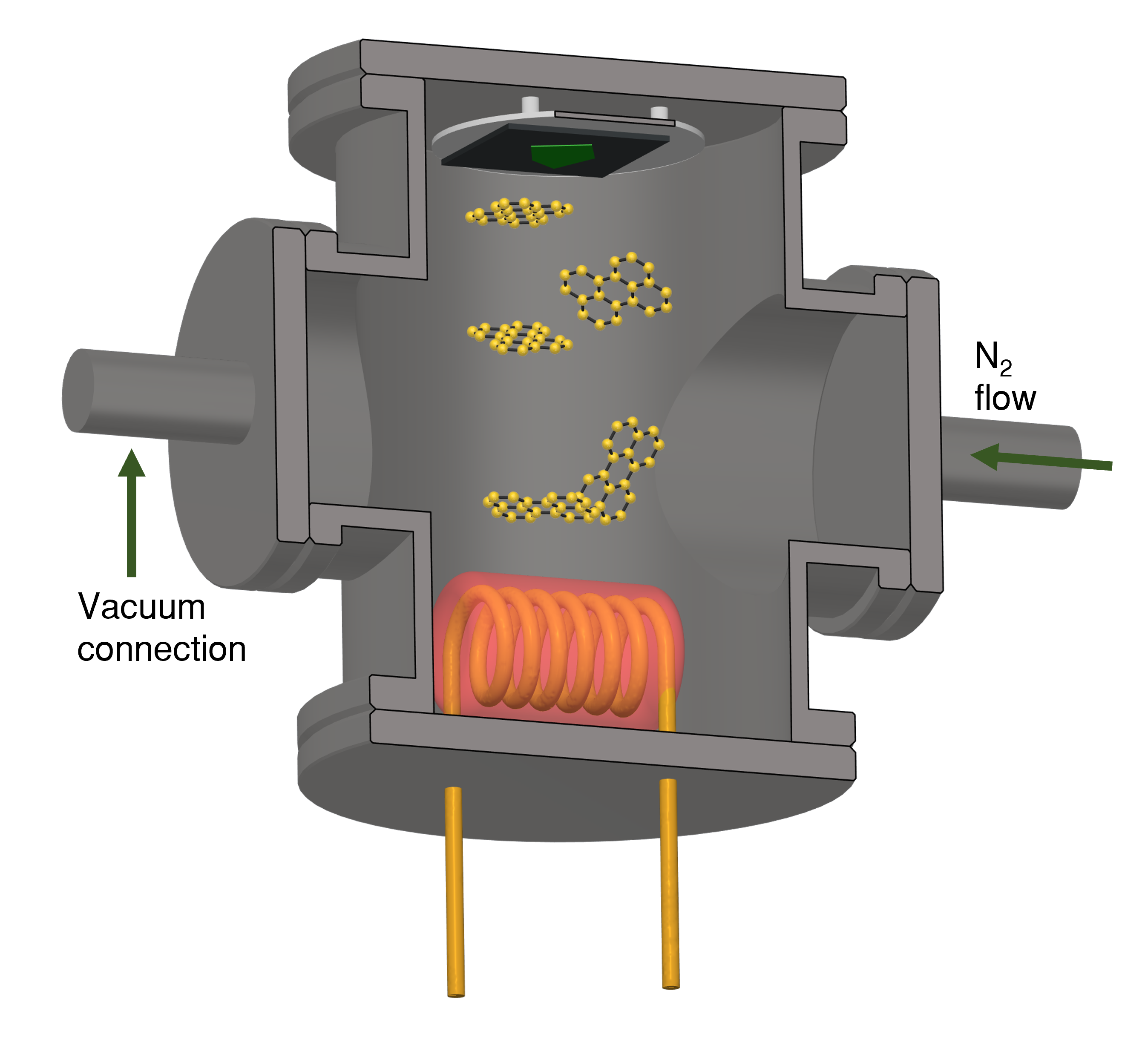}
    \caption{Sketch of the evaporation process inside the chamber. In the top part of the chamber, a substrate holding hBN flakes is mounted onto a chip holder. The vacuum and nitrogen gas connections are equipped on the two sides. The filament at the bottom is being heated, where molecules are evaporated.}
    \label{fig:SI_evapchamber}
\end{figure}

The evaporation of perylene molecules is based on ohmic heating of a resistive nickel-chromium filament inside an evaporation chamber, as illustrated in Fig.\,\ref{fig:SI_evapchamber}. The filament is dipped into a solution of perylene {(Sigma Aldrich, product number: 394475, purity $\geq$ 99.5\%)} in chloroform with a mass ratio of $1:3100$ for $1$ minute. Before each evaporation run, the chamber is repeatedly evacuated and flushed with nitrogen gas.

Afterwards, the chamber is either put under a low vacuum of around 100\,$\mathrm{mbar}$, or under overpressure in nitrogen. {For the latter case, the exact overpressure was not recorded.} In this way, we vary the mean free path of perylene molecules in order to control the deposition yield on the sample. The evaporation times range from a few seconds to one minute for a low vacuum of around 100\,$\mathrm{mbar}$, and many minutes for an overpressure in the chamber (see table \ref{tab:SI_sample_prep_params} for parameters for each sample). For overpressure environments, currents between 600\,$\mathrm{mA}$ and 1\,$\mathrm{A}$ are applied, while for a low vacuum, we use currents between 50\,$\mathrm{mA}$ and 1\,$\mathrm{A}$. The resistances of the different filaments used range from 2\,$\Omega$ to 8\,$\Omega$.

\begin{figure}[h!]
    \centering
    \includegraphics[width=1\linewidth,trim=0cm 0cm 0cm 0cm,clip]{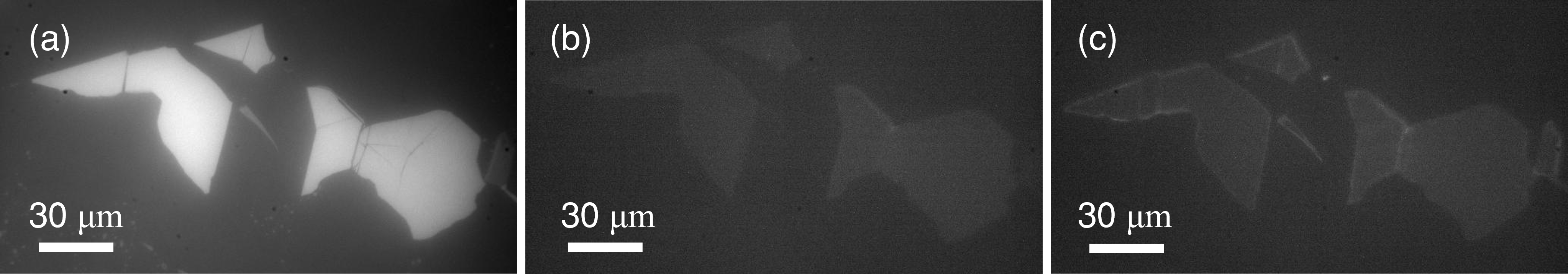}
    \caption{{(a) White-light and (b) fluorescence image of exemplary hBN flakes before evaporation. (c) Fluorescence image of the same flakes directly after an evaporation run. An increase in fluorescence can be observed.}}
    \label{fig:SI_flake_before_after_evap}
\end{figure}

\begin{figure}[h!]
    \centering
    \includegraphics[width=0.9\linewidth]{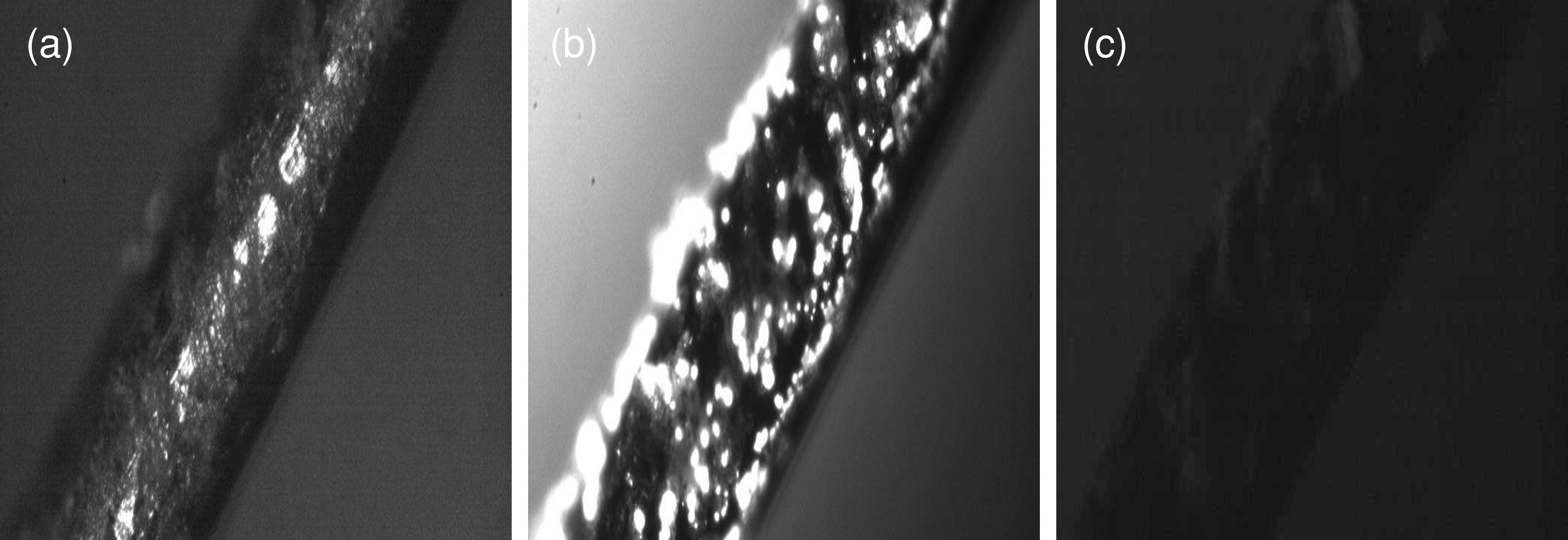}
    \caption{(a) White-light and (b) fluorescence image of a freshly dipped filament before the evaporation run. Fluorescence image of the same area on the filament after the evaporation run is shown in (c). The absence of fluorescence in (c) clearly shows that molecules have been successfully evaporated.}
    \label{fig:SI_wire_evap}
\end{figure}

In Tab.\,\ref{tab:SI_sample_prep_params}, different sample preparation parameters for the corresponding samples are shown. Images of an example hBN flake before and after evaporation are shown in Fig.\,\ref{fig:SI_flake_before_after_evap}. Several optical images of parts of the filaments before and after an evaporation run are presented in Fig.\,\ref{fig:SI_wire_evap}. For Figs.\,\ref{fig:SI_flake_before_after_evap} and \ref{fig:SI_wire_evap}, the observation was done with a custom-built room temperature microscope that utilizes an LED as an excitation source with peak emission at 425\,$\mathrm{nm}$. For fluorescence imaging, 442 nm long-pass and 450 nm long-pass filters were utilized in the emission path.

\newpage

\begin{table}[h!]
\centering
\small
\setlength{\tabcolsep}{4pt}

\begin{tabularx}{\textwidth}{
c
c
c
>{\centering\arraybackslash}X
c
}
\hline
Sample ID &
Evaporation time &
Wire current &
Vacuum level &
Transfer to a pre-cleaned Si-wafer \\
\hline
1 &  $20$\,$\mathrm{min}$ & 1 A & overpressure & no \\
2 &  $1$\,$\mathrm{min}$ & 0.9 A &  $\approx$ 100 mbar & yes \\
3 &  $1$\,$\mathrm{min}$ & 1.07 A & $\approx$ 100 mbar & yes \\
\hline
\end{tabularx}
\caption{Evaporation parameters. For the shown samples in the manuscript, different molecule evaporation parameters, including evaporation time, current through wire, vacuum level during evaporation, are displayed. The pre-cleaned wafers for samples 2 and 3 were first put in an acetone ultrasonic bath for  $5$\,$\mathrm{min}$, then rinsed with ethanol and isopropanol, and finally plasma-etched for  $6$\,$\mathrm{min}$, before transferring the sample.}
\label{tab:SI_sample_prep_params}
\end{table}

\begin{figure}[h!]
    \hspace{0cm}
    \centering    \includegraphics[width=0.8\linewidth,trim=0cm 0.5cm 0cm 0cm,clip]{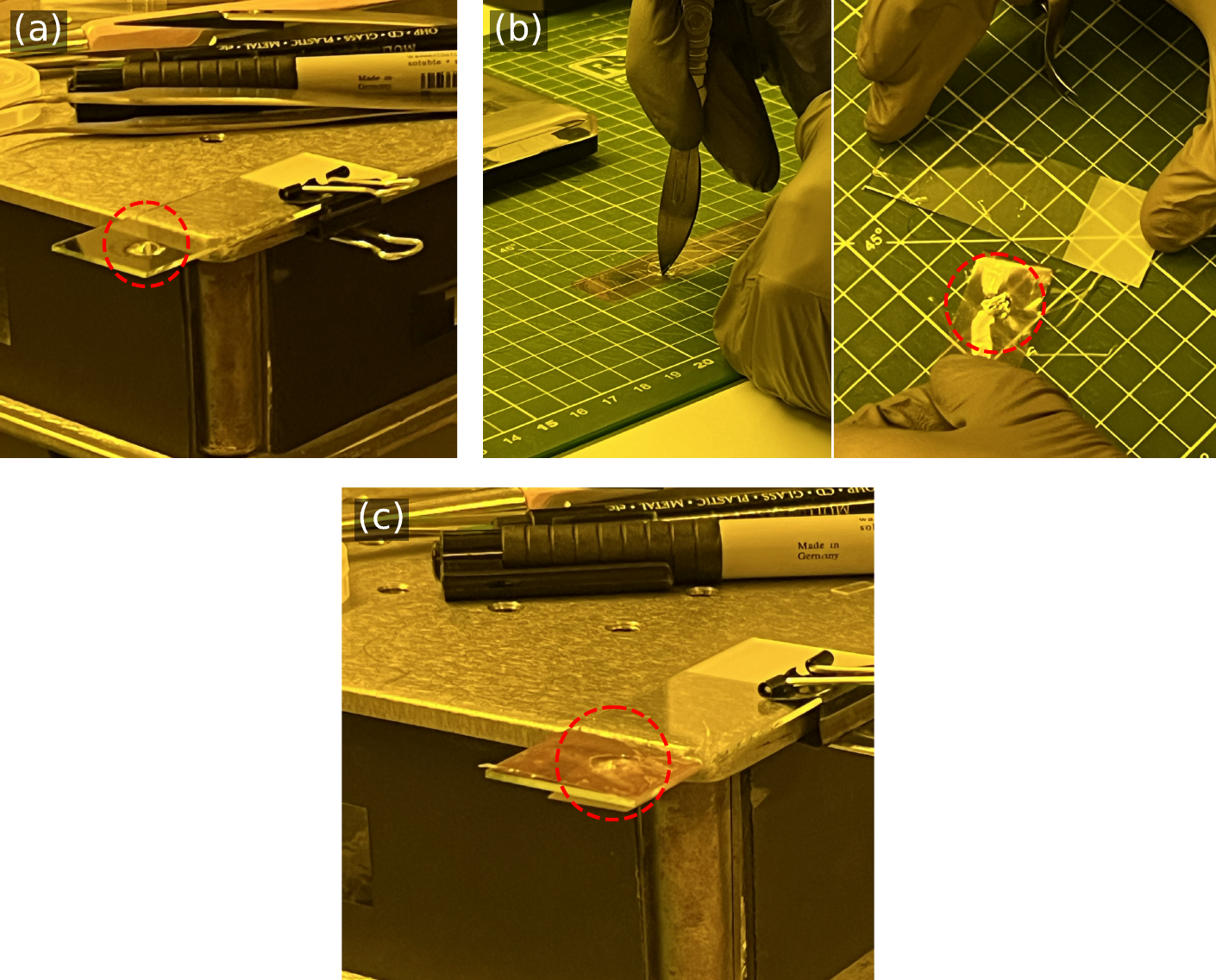}
    \caption{{Stamp fabrication procedure. (a) PDMS dome grown on a microscope {slide}. (b) A circular hole is cut into a piece of Kapton tape with PC (which was spin coated onto another {slide}) transferred onto it afterwards. The hole is now covered with PC. (c) The Kapton tape is put onto the microscope {slide}, such that the PC-covered hole overlaps with the PDMS dome.}}
    \label{fig:SI_stamp_fab}
\end{figure}

\subsection{Fabrication of hBN stacks}\label{hbn_fab}

hBN flakes are first prepared on a clean silicon-wafer by dry-exfoliation\,\cite{Geim_2004} using an adhesive tape (Ultron Systems Inc., product number: 1007R-4.0), and then inspected under an optical microscope. For the exfoliation, a small piece of a bulk hBN crystal (2D Semiconductors, product: BLK-hBN-LG) was transferred onto the adhesive tape and the remaining part was distributed over the whole adhesive area. Then, cut Si-wafer pieces (without pre-cleaning) were placed onto the tape and put onto a hotplate at 50\,°C for $5$\,$\mathrm{min}$. Afterwards, the tape is slowly removed over a period of a few minutes, while hBN flakes remain on the silicon chips. Then, potential candidates are chosen for the bottom and top flakes. The top flake is then picked up at a temperature of 130\,°C with a polydimethylsiloxane (PDMS, product: SYLGARD 184 Silicone Elastomer Base + SYLGARD 184 Silicone Elastomer Curing Agent, 5:1 ratio) dome that is coated with a sacrificial polycarbonate (PC, HQ Graphene, product: 2D\_CL\_PC, 6\% dissolved in chloroform) layer, which together form a ``stamp''. The PDMS domes are fabricated by putting droplets of the Base+Agent mixture onto microscope slides and placing them upside-down into an oven at 80\,°C for $30$\,$\mathrm{min}$. The domes then form under gravity. For PC, slides with thin PC layers are prepared by putting a droplet of the aforementioned PC solution onto them, after which another slide is placed on top, such that the liquid distributes over the whole slide area. Afterwards, the two slides are immediately separated from each other by quickly sliding the top slide away, resulting in a thin PC layer on the remaining slide. 
The subsequent stamp fabrication can be summarized as follows:
\begin{enumerate}[label=\textbf{\arabic*}.]
    \item Pick a microscope {slide} with PDMS domes grown on them [Fig.\,\ref{fig:SI_stamp_fab}(a)].
    \itemsep-0.5em 
    \item Take a piece of Kapton tape, cut a circular hole, and transfer PC {from a microscope slide} onto it, so that the hole is covered with PC [Fig.\,\ref{fig:SI_stamp_fab}(b)].
    \itemsep-0.5em 
    \item Put the Kapton tape onto the PDMS dome, so that the circular hole containing PC aligns with the dome, resulting in a PDMS/PC stamp [Fig.\,\ref{fig:SI_stamp_fab}(c)].
\end{enumerate}

After fabricating the stamp, it is placed onto a hotplate and heated to 165\,°C for $5$\,$\mathrm{min}$. This relaxes the PC layer and results in a smoother surface, increasing the chances of a successful stacking. Then, the silicon chip that holds the bottom flake is put in the evaporation chamber. Perylene molecules are evaporated following the method described in section~1.1.

After evaporation, the silicon chip is mounted onto a copper stage that acts as a thermal conductor. This stage is movable in the lateral direction with micrometer screws and its height can be adjusted using a step motor. The chip is fixed in the middle of the copper stage via a vacuum chuck. The {slide} carrying the stamp is then mounted on its own translational stage and is placed above the silicon chip. The copper stage is then heated to 130\,°C before approaching the stamp. Right before the silicon chip and the stamp come into contact, the stage movement speed is set to the lowest possible speed of 0.14\,$\mathrm{\mu m/s}$, which is then kept during stacking. After the two hBN flakes are properly overlapped, a waiting time of approximately  $10$\,$\mathrm{min}$ is taken to allow the PC layer to relax and thermalize properly with the silicon chip. From here, there are two possible paths of continuation:
\begin{enumerate}[label=\textbf{\arabic*}.]
    \item \textbf{Direct delamination:} The stage is heated to 170\,°C. Right before reaching this temperature, the stage movement speed is set to 0.69\,$\mathrm{\mu m/s}$, and the stage is immediately driven out of contact. During the following delamination, a part of the PC layer detaches from the PDMS dome and remains on the silicon chip. The hBN stack and surrounding hBN flakes are now enveloped by a PC layer.  
    \item \textbf{Delamination on a clean chip after stack pickup:} 
    The whole stack is first picked up and then transferred to a fresh silicon chip, which is cleaned in advance by putting it in an acetone ultrasonic bath for  $5$\,$\mathrm{min}$ with subsequent rinsing with ethanol and isopropanol. After that, it is plasma-etched for  $6$\,$\mathrm{min}$. 
    For the stack pickup, we keep a temperature of 130\,°C. In this case, both the contact and the detachment speed of the stage are kept at 0.14\,$\mu$m/s. The following transfer to the new silicon chip is done as described in point \textbf{1}.
\end{enumerate}

In the general fabrication procedure, the PC layer is removed afterwards by immersing the chip in chloroform (Sigma Aldrich, product number: 32211-1L-M, purity $\geq$ 99.0\%) for 30\,min to 1\,hour, followed by subsequent rinsing with acetone (VWR Chemicals, product number: 20066.321, purity $\geq$ 99.8\%), ethanol (VWR Chemicals, product number: 20821.330, purity $\geq$ 99.8\%) and isopropanol (VWR Chemicals, product number: 20880.320, purity $\geq$ 99.8\%). 

\begin{figure}[h!]
    \centering
    \includegraphics[width=0.9\linewidth]{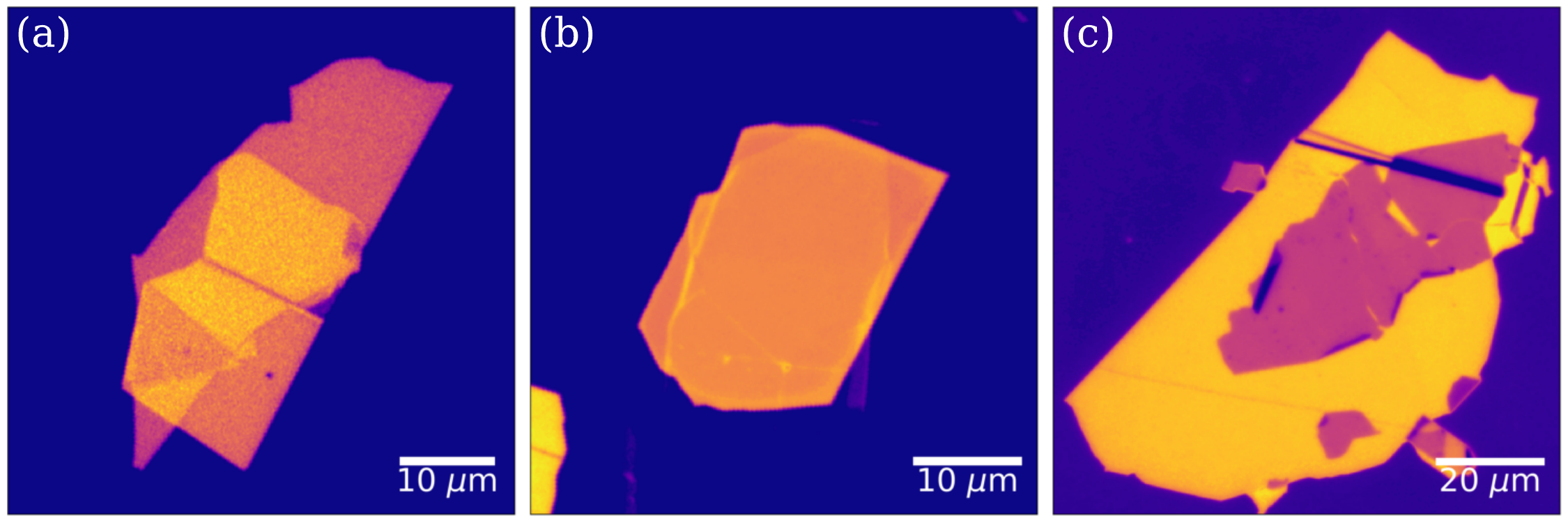}
    \caption{Overview of the three fabricated perylene/hBN samples investigated in this work. (a), (b), (c) show optical microscope images of Samples 1, 2, and 3, respectively.}
    \label{fig:SI_samples_overview}
\end{figure}

Figure\,\ref{fig:SI_samples_overview} shows optical microscope images of three samples used for measurements reported in the manuscript.

\subsection{Fluorescence images}

\begin{figure}[h!]
    \centering
    \includegraphics[width=0.9\linewidth,trim=0cm 0cm 0cm 0cm,clip]{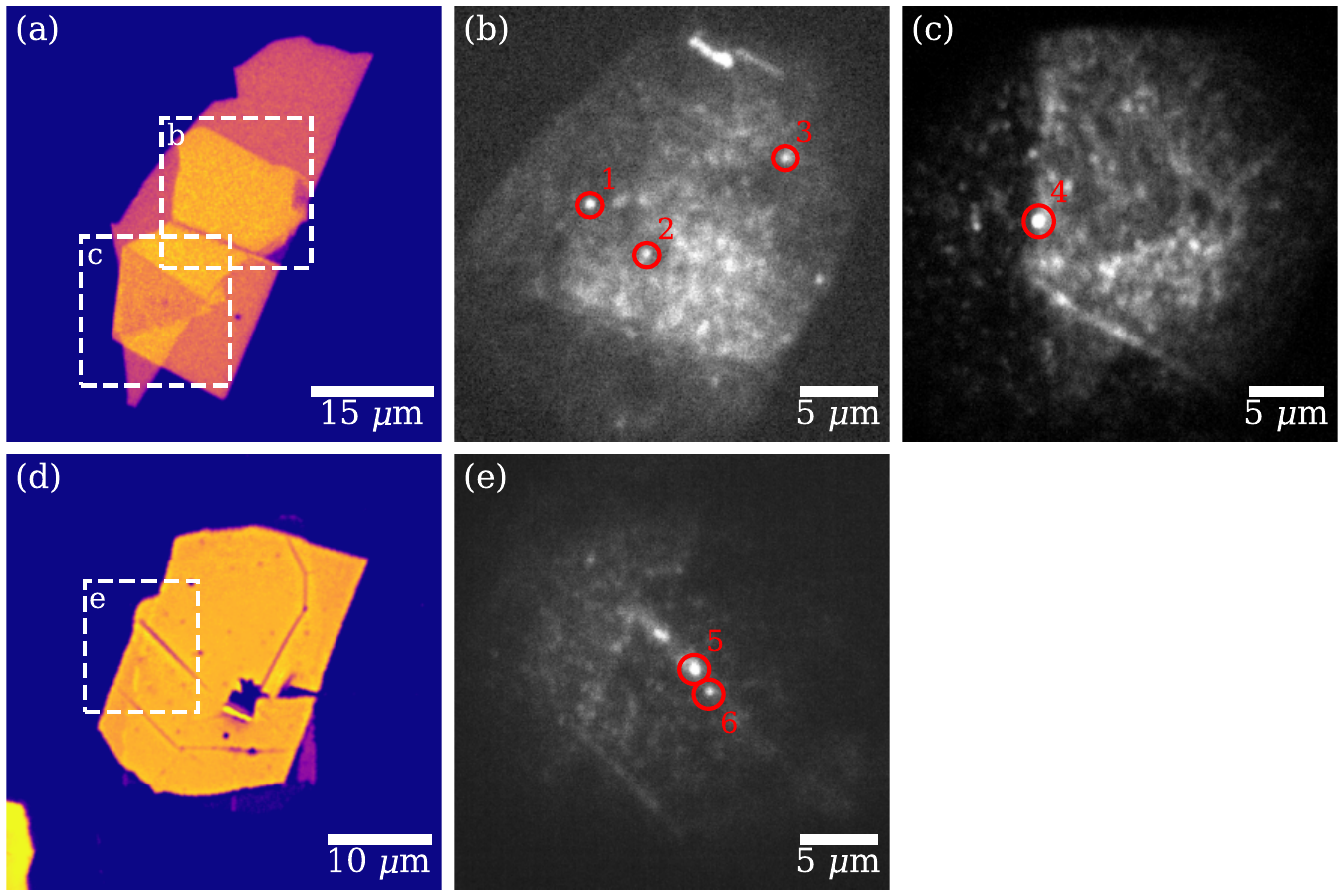}
    \caption{(a), (d) Overview microscope images of the investigated perylene/hBN samples. The white dashed boxes indicate the field of view shown in the corresponding enlarged fluorescence maps. (b), (c), (e) Fluorescence maps of the selected regions. The red circles indicate the individual molecular sites from which the spectral data were acquired.}
    \label{fig:SI_target_map}
\end{figure}

Here, we relate the spectral data presented in the manuscript to the spatial locations of the corresponding emission sites on the respective samples. Fluorescence maps of the investigated samples are shown in Fig.\,\ref{fig:SI_target_map}. The overview images in Figs.\,\ref{fig:SI_target_map}(a),(d) correspond to Sample 1 and Sample 2, respectively. The selected regions indicate the sample areas from which the emission spectra, spectral diffusion traces, and resonance-excitation data shown in Figs.\,2--4 of the manuscript were acquired. The sites of individual molecules are directly labeled in the fluorescence maps with red circles, and their correspondence to the data presented in the manuscript is summarized in Tab.\,\ref{tab:target_map}.

\begin{table}[h!]
\centering
\small
\setlength{\tabcolsep}{8pt}
\begin{tabular}{lc}
\hline
\multicolumn{1}{c}{Figure} & Fluorescence Map \\
\hline
\hspace{6mm}Fig. 2(a),(c)-i & (e)-5 \\
\hspace{6mm}Fig. 2(c)-ii & (b)-1 \\
\hspace{6mm}Fig. 2(c)-iii & (e)-6 \\
\hspace{6mm}Fig. 2(c)-iv & (b)-3 \\
\hspace{6mm}Fig. 3(a) & (b)-2 \\
\hspace{6mm}Fig. 3(c) & (e)-5 \\
\hspace{6mm}Fig. 4 & (c)-4 \\
\hline
\end{tabular}
\caption{{Correspondence between the spectral data shown in Figs. 2–4 of the manuscript and the measurement positions labeled in Fig.\,\ref{fig:SI_target_map}}}
\label{tab:target_map}
\end{table}

\subsection{Perylene in anthracene sample fabrication}

The sample was prepared through sublimation of a mixture of perylene (Sigma Aldrich product number P11204) and anthracene (Sigma Aldrich, product number 10580, purity > 99.0\%). The anthracene material was additionally purified in a self-built zone refiner. The mixture had a mass ratio of perylene to anthracene of $1 : 2\,700\,000$.

A mass of 9.7\,$\mathrm{mg}$ of mixture was put into a plasma-cleaned flask. The flask was evacuated and subsequently filled with around 1\,$\mathrm{atm}$ of argon gas. A heat gun set to $400\,^{\circ}\mathrm{C}$ was pointed underneath the flask, sublimating the mixture. Within three minutes crystals appeared in the flask. The flask was then emptied onto a petri dish from which a high-quality crystal was selected.

\newpage

\section{Experimental setup}

\begin{figure}[h!]
    \centering
    \includegraphics[width=0.6\linewidth, trim=0cm 0.1cm 0cm 0.3cm,clip]{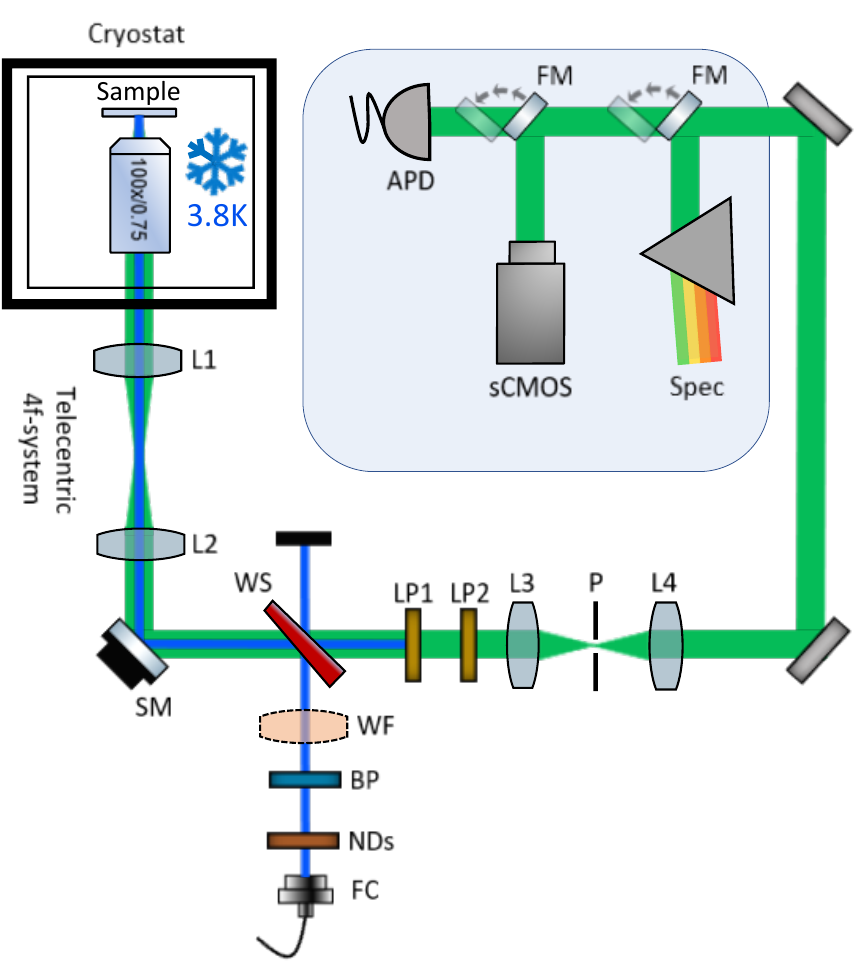}
    \caption{Sketch of the experimental setup. \textbf{L1}--\textbf{L4}: lenses. \textbf{BP} and \textbf{LP1}--\textbf{LP2}: bandpass and longpass filters. \textbf{SM}: steering mirror. \textbf{WF}: wide-field lens. \textbf{WS}: wedged sampler. \textbf{FC}: fiber collimator. \textbf{FM}: flip mirror.   \textbf{sCMOS}: scientific CMOS camera. \textbf{Spec}: spectrometer. \textbf{APD}: avalanche photodiode. \textbf{P}: confocal pinhole. \textbf{NDs}: neutral-density filters.}
    \label{fig:SI_expsetup}
\end{figure} 

The experimental setup sketched in Fig.\,\ref{fig:SI_expsetup} allows us to perform single-molecule microscopy and spectroscopy at cryogenic temperatures below 4\,K (Montana Instruments, CryoAdvance 100). For the illumination path, the excitation laser, which can be a diode laser at 454\,nm (Toptica Photonics, DL pro HP 454), a diode laser at 423\,nm (Omicron-Laser, LuxX 425--120), or the second-harmonic component of a Ti:Sapphire laser from 440\,nm to 480\,nm (Sirah Lasertechnik, Matisse CS--WaveTrain), is delivered through a single-mode fiber. The laser power can be attenuated in discrete steps using neutral-density filters. Afterwards, the laser passes through a bandpass filter to clean its spectral profile, and then it encounters a wedged sampler, where 90\% of the beam is transmitted and 10\% is reflected. The reflected beam then propagates through a telecentric imaging system constructed by a 4f-lens system together with a steering mirror, enabling scan of the laser focus. After entering the cryostat, the laser is focused on the sample via an objective with a numerical aperture of 0.75 (Zeiss, LD EC Epiplan--Neofluar 100x/0.75 DIC M27). Furthermore, a wide-field lens can optionally be placed in front of the wedged sampler, allowing wide-field fluorescence imaging. 

Fluorescence photons from the sample are collected by the same objective and guided backward through the common path before being directed to the wedged sampler, where 90\% of photons are transmitted towards a set of longpass filters that filter out the excitation laser. The remaining fluorescence signal is focused through a confocal pinhole to reject out-of-focus fluorescence. Finally, signals are guided to various detection devices, including a low-noise sCMOS (Teledyne, Prime BSI), a spectrometer (Teledyne, SpectraPro HRS--300--S), and other detectors, depending on the measurement method. 

\newpage

\section{Numerical calculations}\label{ab_initio}

The numerical study in this work involves three components. First, (time-dependent) density-functional theory (TDDFT) calculations are carried out using the ORCA software package\,\cite{SI-neese2025}. The equilibrium ground-state geometry is obtained through structural relaxation employing the B3LYP exchange–correlation functional\,\cite{SI-lee1988development,SI-becke1993density}, augmented by D4 dispersion corrections\,\cite{Caldeweyher2019} and the def-TZVPD basis set. Starting from these optimized geometries, we compute excited-state properties using Casida's linear-response time-dependent DFT formalism~\cite{SI-casida1996time}, including the Tamm–Dancoff approximation. The (TD)DFT calculations provide access to the relaxed molecular structure in the $S_0$ and $S_1$ manifolds that are required to estimate the Huang-Rhys factors.

The second step addresses the phononic structure of the extended host. Our simulations address configurations in which perylene is incorporated either between extended AA'-stacked hBN slabs that deform in response to the molecule, or at extended AA'-stacked defective structures, such as edges and steps. Density-functional theory becomes costly for such systems that involve thousands of atoms. Instead, we model molecular motion using the ASE software package\,\cite{SI-hjorth2017atomic} in combination with the foundation machine-learning interatomic potential MACE-MP (medium size)\,\cite{batatia2023foundation2} and explicitly account for dispersive interactions. 

Thirdly, we utilize MACE-MP in combination with structure relaxation to gain a microscopic understanding of the possible binding-sites of perylene. All cell parameters and nuclear positions are relaxed for $50\,000$\,steps using the FIRE optimization algorithm.

\subsection{Visualization of insertion geometries}\label{visualization}

\begin{figure}[h!]
    \centering
    \includegraphics[width=1\linewidth]{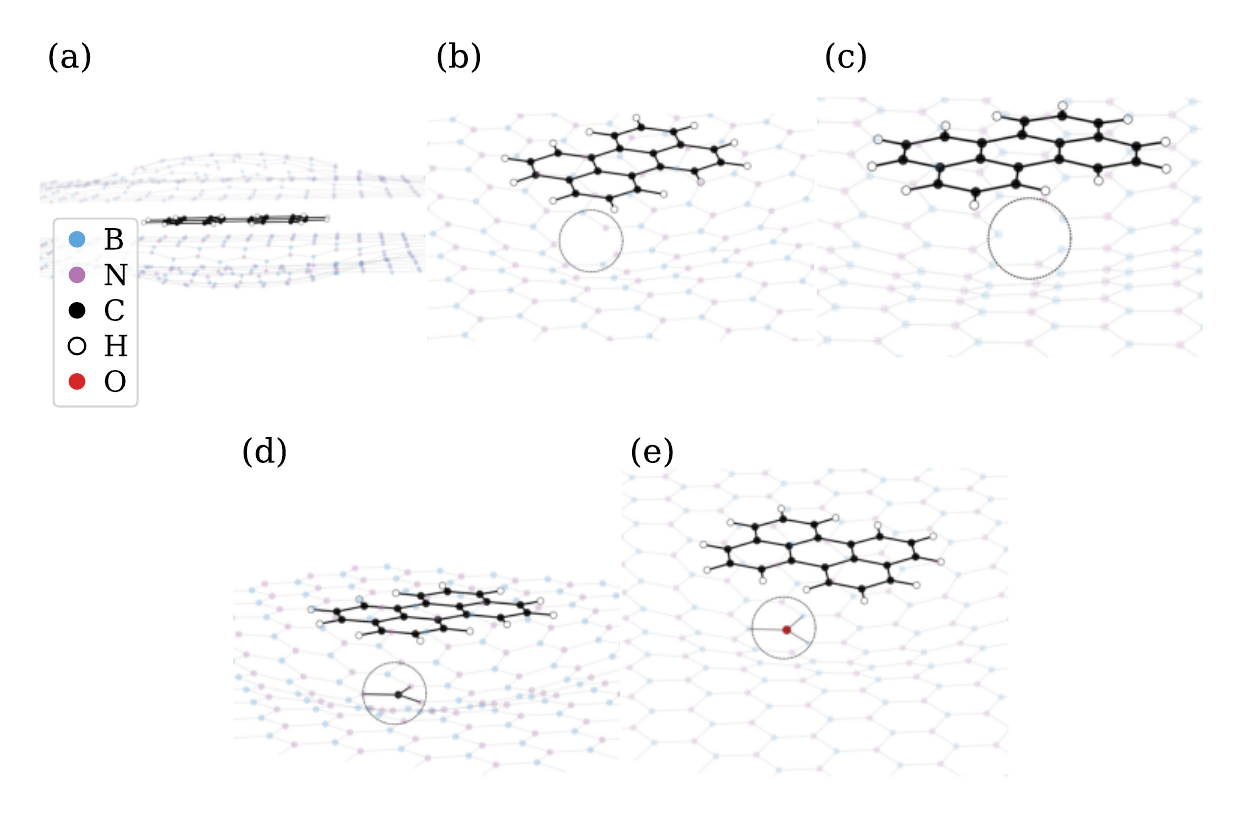}
    \caption{Perylene in an hBN bilayer stack (E1--E5). (a) The side view for perylene in the pristine structure. This side view is identical for the other structures in this figure as well. (b-e) Close-ups into the inside of the stack, with different visible {substitutional} point defects (inside black circles with dashed lines), where (b) and (c) show B- and N-vacancies, respectively. For these images, the top layer has been omitted for visualization purposes.}
    \label{bilayer}
\end{figure}

\begin{figure}[h!]
    \centering
    \includegraphics[width=1\linewidth]{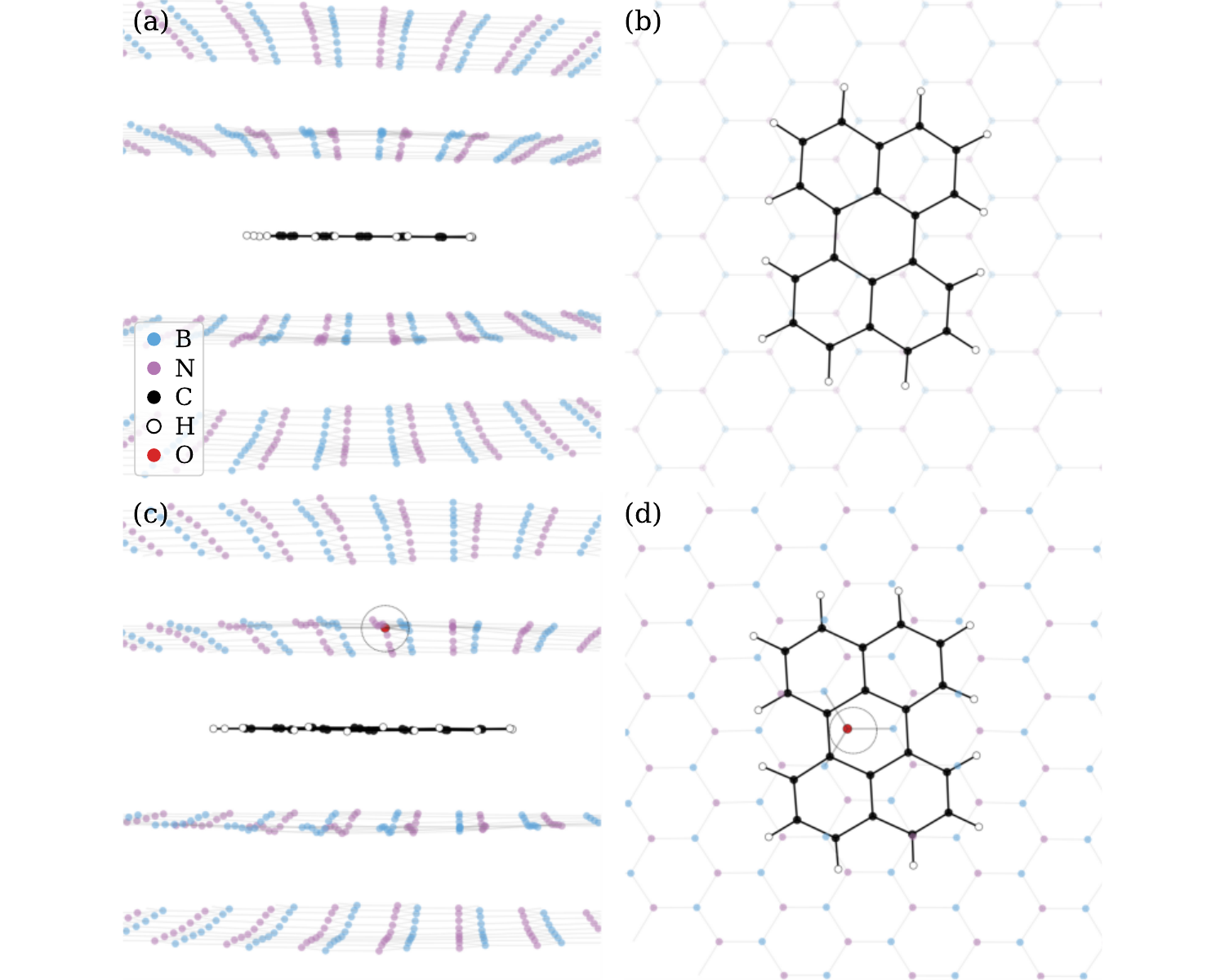}
    \caption{Perylene in an hBN multilayer stack (E6--E7). (a), (b) Perylene side and top views in the pristine structure. (c), (d) Perylene side and bottom views in an hBN structure containing an oxygen point defect. For the top views in (b) and (c), most top and bottom layers have been omitted for visualization purposes.}
    \label{bislab}
\end{figure}

\begin{figure}[h!]
    \centering
    \includegraphics[width=1.1\linewidth]{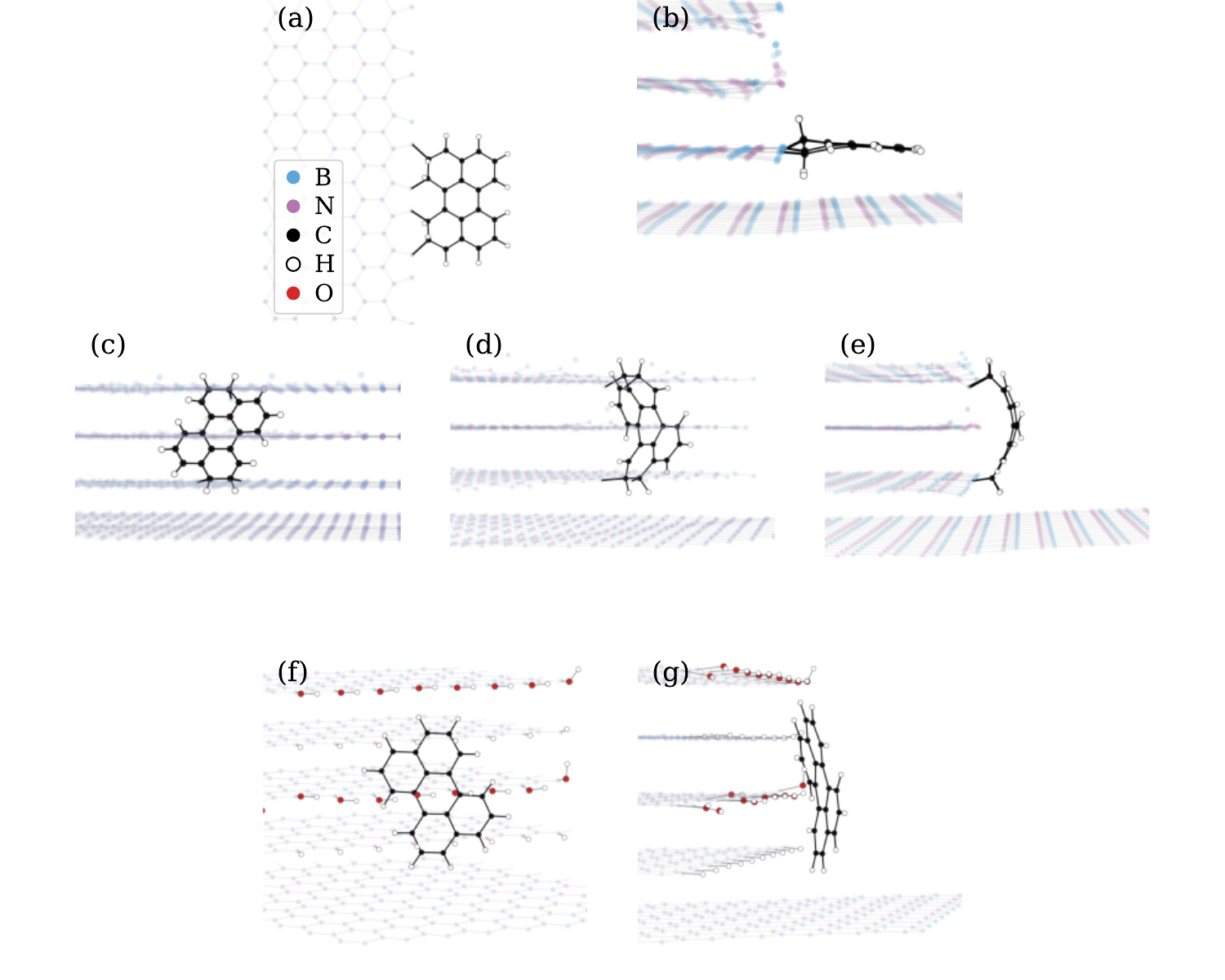}
    \caption{Perylene in an hBN multi-step edge structure (E8--E10). (a), (b) Perylene ( ``lying'') top and side views in the pristine structure. One can see B and N atoms forming covalent interlayer bonds in some parts of the edge. (c-e) Perylene (``standing'') front and side views in the pristine structure. (f), (g) Perylene (``standing'') front and side views in an OH/H-terminated structure. For the top view in (a), most layers have been omitted for visualization purposes. Structures E8 and E9 contain covalent bonds between the molecules' C atoms and B atoms.}
    \label{bislabedge}
\end{figure}

\begin{figure}[h!]
    \centering
    \includegraphics[width=0.7\linewidth]{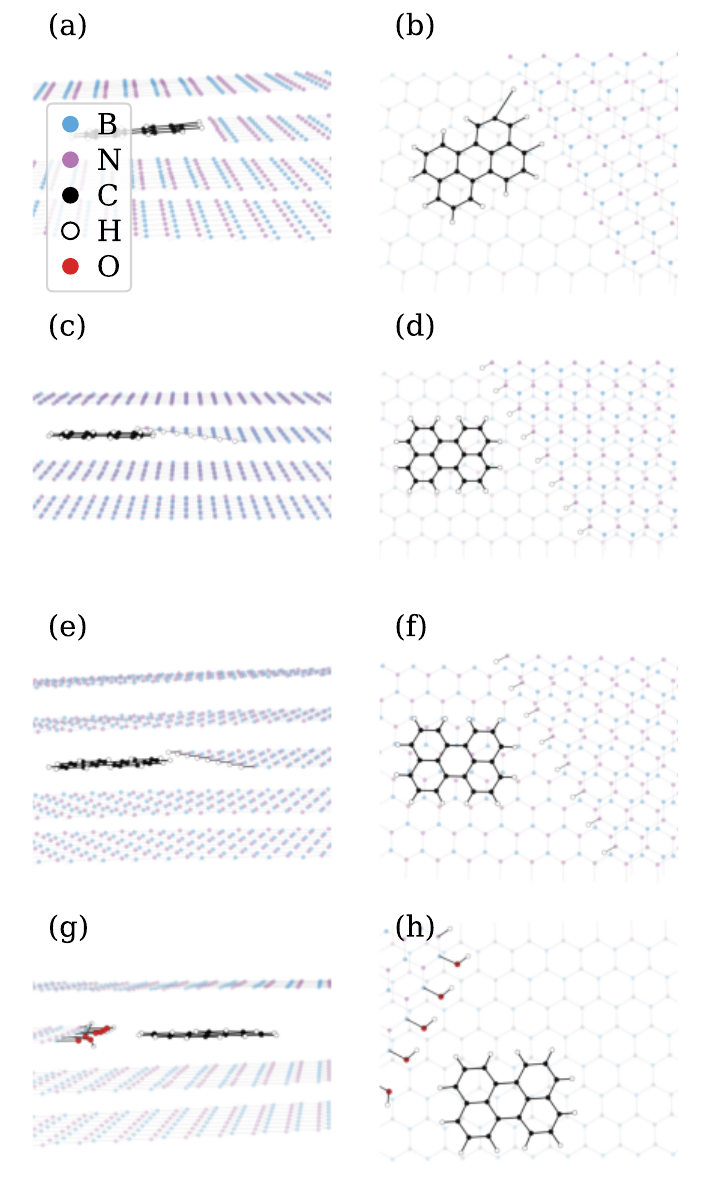}
    \caption{Perylene in an hBN embedded step edge structure (E11--E14). (a), (b) Perylene side and top views in the pristine structure. (c), (d) Perylene side and top views in a H-terminated structure. (e), (f) Perylene side and top views in a larger H-terminated structure, containing an additional hBN layer. (g), (h) Perylene side and top views in an OH-terminated structure. For the top views, most layers have been omitted for visualization purposes.}
    \label{embeddedstep}
\end{figure}

\begin{figure}[h!]
    \centering
    \includegraphics[width=1\linewidth]{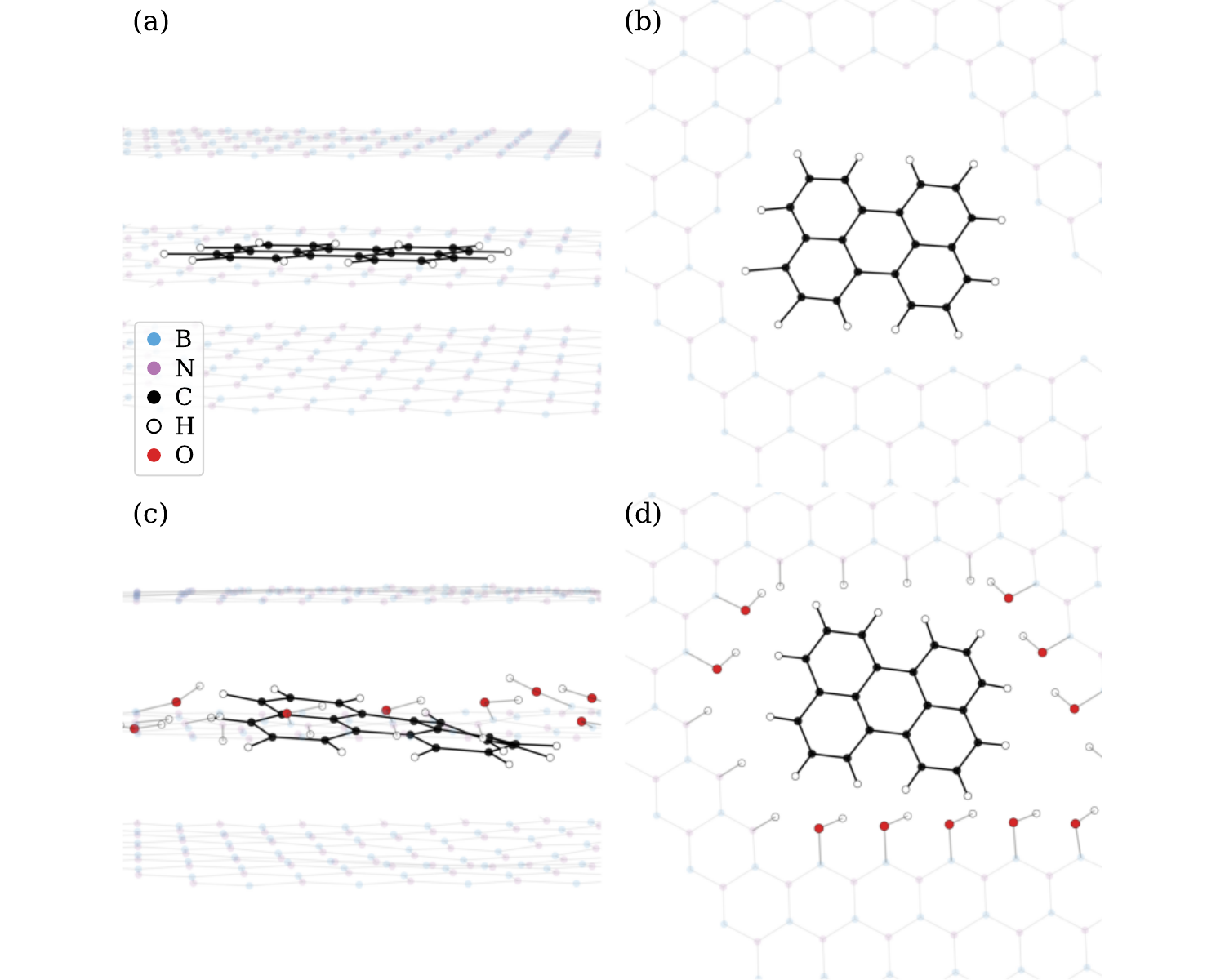}
    \caption{Perylene in an hBN extended vacancy structure (E15--E16). (a), (b) Perylene side and top views in an unterminated vacancy structure. (c), (d) Perylene side and top views in a H- and OH-terminated vacancy structure. For the top views, most layers have been omitted for visualization purposes.}
    \label{extendedvacancy}
\end{figure}

In this section, we plot the insertion geometries of all numerical calculations presented in Tab.\,2 of the manuscript. The calculations were performed for a single perylene molecule embedded in various hBN structures. For each structure, the formation energy is calculated as $E_f = E_\mathrm{Mol + hBN} - (E_\mathrm{Mol} + E_\mathrm{hBN})$, where $E_\mathrm{Mol + hBN}$ is the total energy of the combined molecule-hBN system, and $E_\mathrm{Mol}$ and $E_\mathrm{hBN}$ are the energies evaluated with MACE-MP of separately relaxed molecule and hBN structure, respectively. The formation energy therefore describes how stable the formation of a bound perylene-hBN system is. The corresponding structures E1 - E16 in Tab.\,2 of the manuscript are shown in the Figs.\,\ref{bilayer} - \ref{extendedvacancy}. For views from the top, several layers have been omitted for visualization purposes.

\clearpage

\newpage

\newpage

\clearpage

\subsection{Diffusion barrier of perylene in an hBN stack}\label{diffusion_perylene}

\begin{figure}[h!]
    \centering
    \includegraphics[width=0.8\linewidth]{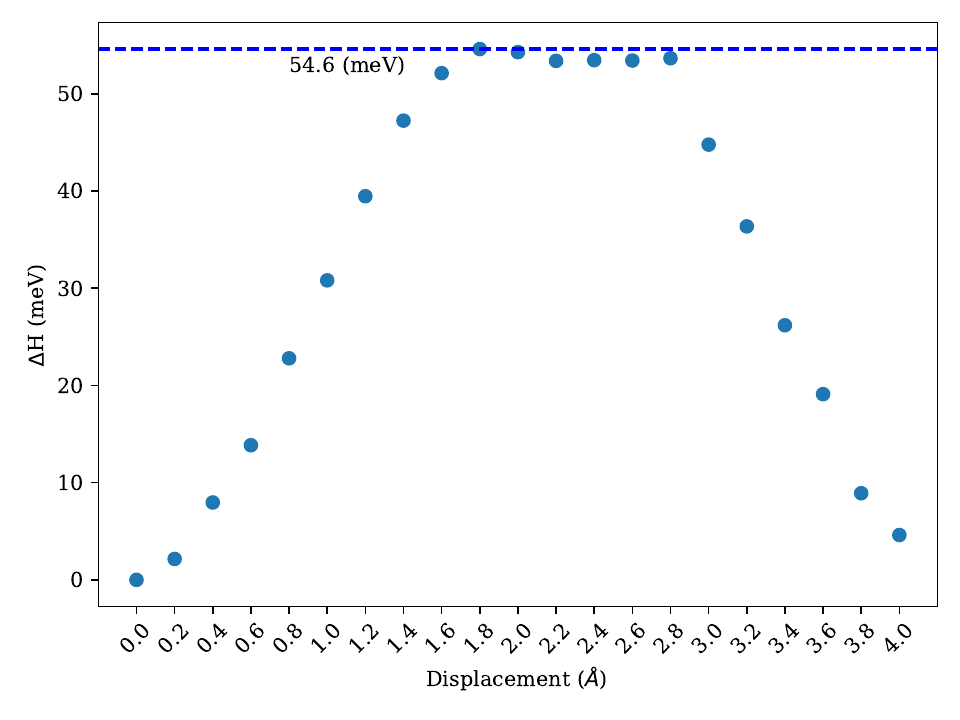}
    \caption{Relative enthalpy of perylene in an hBN stack versus the displacement of perylene from one lattice site to another. It can be seen that the diffusion barrier is around 54.6\,meV. After reaching the next lattice site, the relative enthalpy decreases again, until it reaches the next minimum.}
    \label{fig:SI_perylene_diffusion}
\end{figure}

Fig.\,\ref{fig:SI_perylene_diffusion} displays the relative enthalpy of the perylene/hBN system when the molecule is displaced by up to one hBN lattice in the stack. The geometry of the molecule is held fixed while the hBN stack is allowed to relax after each step. It can be seen that a plateau is reached after a displacement of around 1.8 Å, which acts as an upper bound in the energy needed to displace the perylene molecule from one lattice site to the
next. Since this energy is comparable to the thermal energy at room temperature (25.7\,meV at 300\,K), the associated Boltzmann factor is therefore $\exp(-54.6 \,\text{meV}/ 25.7 \,\text{meV}) \approx 0.12$, which indicates that thermally activated hopping between neighbouring hBN lattice sites is plausible at room temperature. The hopping yield would further increase during the stacking and delamination process, when the stack is heated up to around 170\,°C, and the resulting Boltzmann factor is $\exp(-54.6 \,\text{meV}/ 38.2 \,\text{meV}) \approx 0.24$. In addition, the upper stack is approached onto the lower flake at an inclined angle. This geometrical factor should also assist perylene molecules to be progressively expelled from the pristine stack.

\newpage

\section{Emission spectra}

In this section, we present the Franck-Condon simulation algorithm performed for perylene in hBN, together with the phonon-broadening model used to account for the coupling to lattice vibrations. The resulting vibronic spectra are then compared with the experimentally measured emission spectra of perylene in hBN, as shown in Fig.\,2(c) of the manuscript.

\subsection{Calculation of Franck-Condon factors}

Starting from relaxed ground- and excited-state geometries, vibronic spectra were calculated within the displaced harmonic oscillator approximation. The electronic ground and excited potential energy surfaces are assumed to be harmonic and to share identical normal modes.

Normal modes were expressed in mass-weighted Cartesian coordinates. The displacement between the ground- and excited-state equilibrium geometries of the molecule, $\Delta \mathbf{R}$, is projected onto each mass-weighted normal mode $\mathbf{e}_{ij}$ (where $i$ is an index describing the number of atoms with their $3$ degrees of freedom and $j$ describes the number of all modes) to obtain the normal-coordinate displacements,
\begin{equation}
\Delta Q_j = \sum_i \sqrt{m_i}\,\mathbf{e}_{ij}\cdot \Delta\mathbf{R}_i .
\end{equation}
Here, $m_i$ denotes the atomic masses, and the index $i$ in this case runs over all perylene atoms (and their respective $3$ degrees of freedom, so $96$ in total). For our simulations, the resulting $\Delta Q_j$ is now a vector with length equal to the total number of modes (coming from all atoms in the hBN-structures $+$ $32$ perylene atoms, each with their $3$ degrees of freedom). From this, we calculate the Huang-Rhys factor for each mode $j$ \cite{Alkauskas_2014}.
\begin{equation}
S_j = \frac{\omega_j}{2\hbar} (\Delta Q_j)^2 ,
\end{equation}
with $\omega_j = 2\pi c\,\nu_j$, where $\nu_j$ is the vibrational wavenumber. 

Within the harmonic approximation, the Franck-Condon (FC) overlap between the vibrational ground state and the $v_m$-th excited state of mode $m$ follows a Poisson distribution \cite{Wen},
\begin{equation}
P_m(v_m) = e^{-S_m}\,\frac{S_m^{v_m}}{v_m!}.
\end{equation}
In order to decrease the calculation time, a threshold for the FC-overlaps is set, in order to filter out modes that do not have any significant contribution. With the resulting ``active modes'', the rest of the simulation was performed.

Multimode vibronic transitions were constructed by enumerating all combinations of vibrational quantum numbers $\{v_m\}$ up to a maximum total vibrational excitation $v_{\mathrm{max}}$ (with $v_{\mathrm{max}} \geq \sum_m v_m$). Each transition contributes with a wavenumber shift
\begin{equation}
\Delta \nu = \sum_m v_m \nu_m
\end{equation}
with a corresponding Franck-Condon weight
\begin{equation}
W(\{v_m\}) = \prod_m P_m(v_m).
\end{equation}
This procedure creates a discrete spectrum, as seen in Fig.\,\ref{fig:overtone}(a). For this specific calculation, only 6 active modes contribute to the spectrum. The occupation quantum numbers for each peak numbers are shown on the top axis of the figure.

\subsection{Phonon broadening model}

After calculating the Franck-Condon weights for the respective transitions, a broadening model was used to model phonon broadening of each Franck-Condon transition. Simulation peaks that were very close to each other (distance of up to 5 cm$^{-1}$) were binned together, so that many energetically close peaks are replaced by one stronger peak.

The model assumes that every peak can be split into a narrow zero-phonon contribution (linewidth limited by the spectrometer resolution) and a broader phonon-sideband contribution due to Debye-Waller coupling, inspired by Grushka\cite{Grushka1970}:
\begin{equation}
S(\nu)=\sum_k w_k\left[\alpha_{\mathrm{DW}}\,G(\nu;\nu_k,\sigma_k) + (1-\alpha_{\mathrm{DW}})\,\mathrm{EMG}(\nu;\nu_k,\sigma_k,\tau_{\mathrm{ph}})\right],
\end{equation}
where $G$ is a unit-area Gaussian and $\mathrm{EMG}$ is a unit-area exponentially modified Gaussian (Gaussian convolved with a one-sided exponential tail with decay length $\tau_{\mathrm{ph}} = 70$\,cm$^{-1}$). $w_k$ are the weights of the corresponding Franck-Condon bars and $\alpha_\mathrm{DW}$ is the mean Debye-Waller factor obtained from the measured spectra using $\alpha_\mathrm{DW} = I_\mathrm{ZPL}/(I_\mathrm{ZPL} + I_\mathrm{Phonon})$. Here, we determine a mean value of $\alpha_\mathrm{DW} = {0.49 \pm 0.01}$. For $\sigma{\mathrm{ph}}$, a global ZPL linewidth of $9.5$\,cm$^{-1}$ is used for every peak, which, for our emission spectra, is limited by the resolution of the spectrometer. This global linewidth is obtained by fitting a Gaussian to the 00-ZPL.

\begin{figure}[h!]
    \centering
    \includegraphics[angle=270, width=0.99\linewidth]{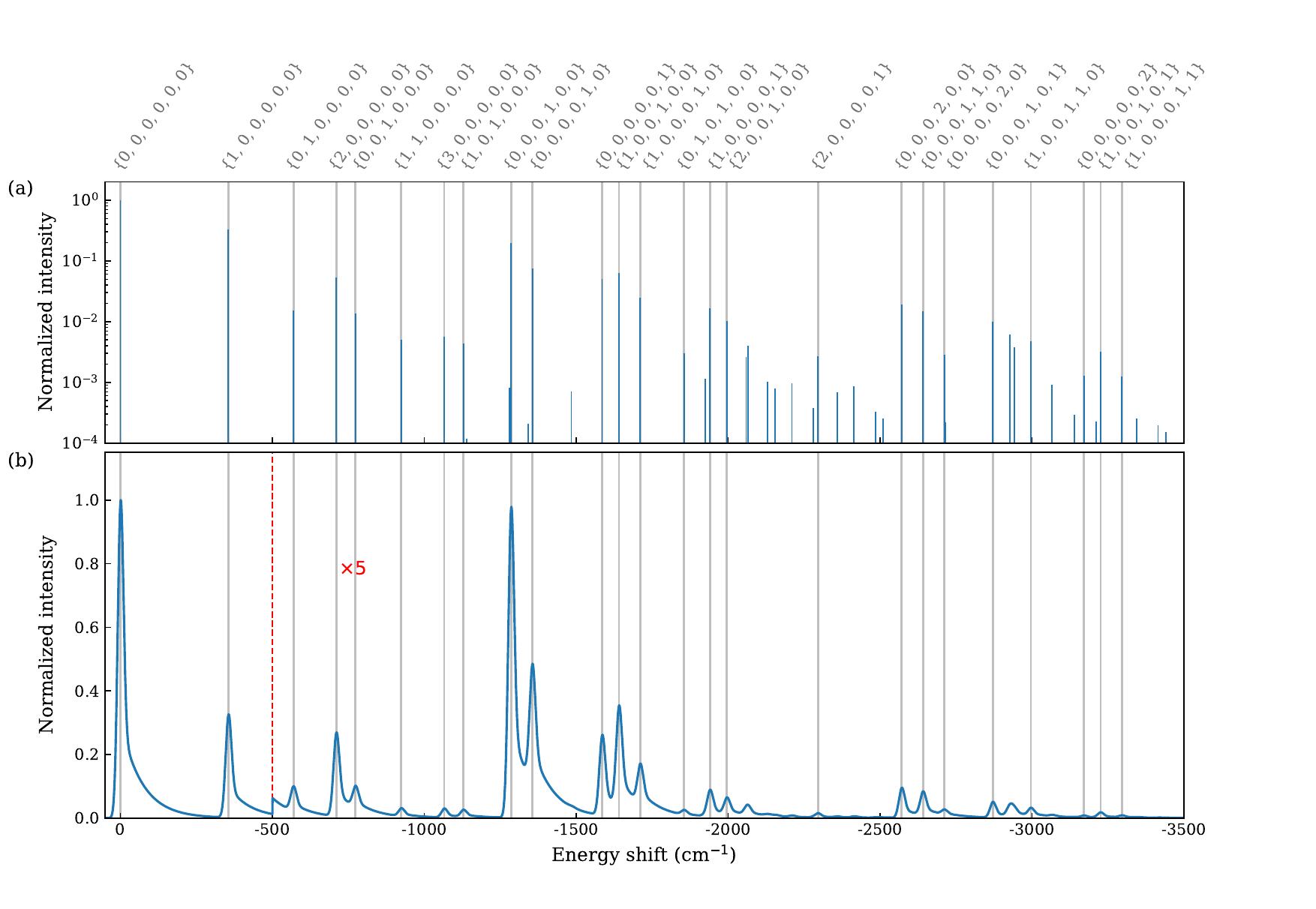}
    \vspace{-20mm}
    \caption{(a) Franck-Condon simulated spectrum. (b) Phonon-broadening model applied to the spectrum in (a).}
    \label{fig:overtone}
\end{figure}

In Fig.\,\ref{fig:overtone}(b), a broadened spectrum can be seen. Here, the discrete spectrum in Fig.\,\ref{fig:overtone}(a) is used for the broadening procedure. The red dashed line shows the point after which the whole broadened spectrum has been scaled up by a factor for visualization.

\subsection{Temperature-dependent emission spectra}

\begin{figure}[h!]
    \centering    \includegraphics[width=0.9\linewidth,trim=0cm 0.6cm 0cm 0.6cm,clip]{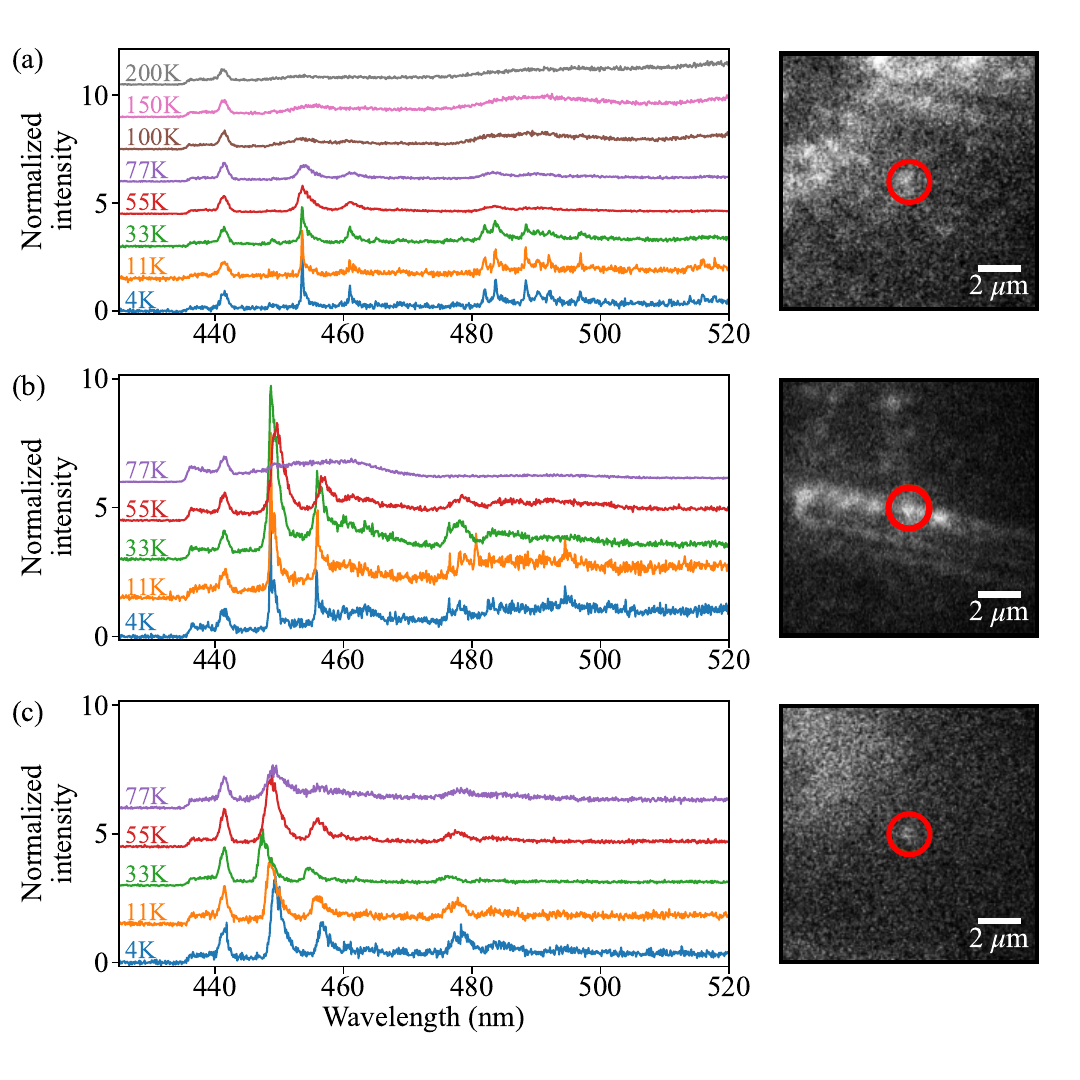}
    \caption{Temperature-dependent fluorescence spectra. (a)–(c) Spectra of individual perylene emission sites in the hBN stack recorded at various temperatures, showing thermal broadening of the spectral features. The right panels show the corresponding fluorescence images, with the measured emitters marked by red circles. All spectra are vertically offset for clarity.}
    \label{fig:SI_high_T}
\end{figure}

Fig.\,\ref{fig:SI_high_T} shows the emission spectra of three perylene sites at different temperatures. As the temperature increases from 4\,K, phonon coupling becomes more pronounced, and the initially narrow ZPL and vibronic peaks gradually broaden and lose contrast. Starting around 55\,K, the ZPLs are overwhelmed by their respective phonon sidebands. The peak between 440 and 443\,nm maintains a constant profile, regardless of measurement position on the sample. The amount of red shift from the excitation wavelength 423.3\,nm, which is $900$ to $1000\,\mathrm{cm}^{-1}$, matches the second-order Raman scattering from silicon.\cite{parker1967raman} The first-order Raman scattering expected at $520\,\mathrm{cm}^{-1}$ is blocked by the longpass filter.

Several experimental factors should be considered when interpreting this series of measurements. On the one hand, when the temperature is increased, not only does the linewidth of perylene molecules broaden, but the transition peaks of other organic residues also broaden. Thus, they are more likely to be excited to fluoresce at higher temperatures, affecting the distinguishability of the target molecules and limiting further measurements. On the other hand, photobleaching occurs more readily at higher temperatures, especially when the temperature approaches 77\,K. In addition, the temperature changes cause spatial drift of the sample, as the sample is mounted on a nano-positioner. Consequently, the lateral position and focal length of the objective need to be readjusted after each temperature change, introducing variations in the illumination and photon collection conditions.

\clearpage

\section{Spectral diffusion data analysis}

\begin{figure}[h!]
    \centering    \includegraphics[width=0.9\linewidth,trim=0cm 0.2cm 0cm 0.4cm,clip]{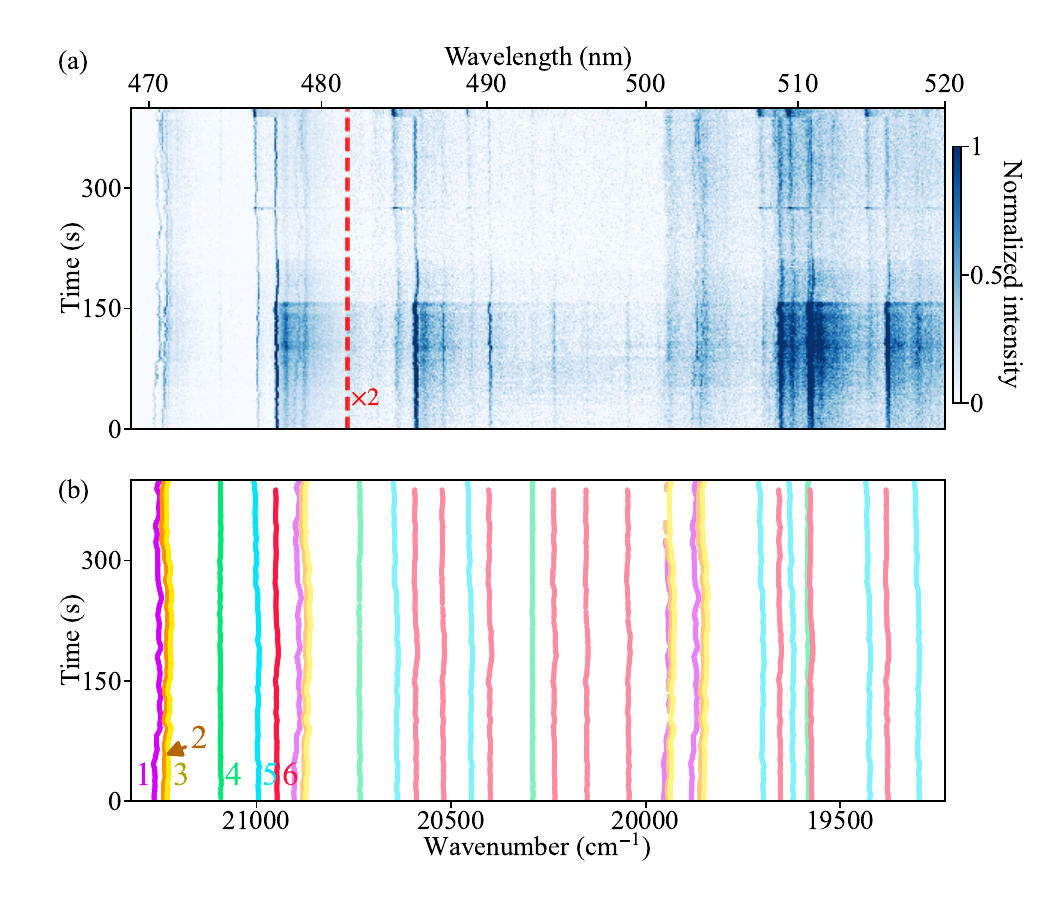}
    \caption{Spectral evolution of six molecules within one diffraction-limited spot. (a) Time-resolved emission spectra recorded over 400 s. (b) Extracted spectral diffusion traces of the identifiable ZPL and vibronic peaks of six molecules. Peaks belonging to the same molecule are labeled by the same color. The bold lines denote the ZPL transitions, and the narrow lines are the vibronic transitions.}
    \label{fig:SI_spectral_diffusion}
\end{figure}

In this section, we provide more detailed analysis and discussion of the spectra presented in Fig.\,3(a) of the manuscript. The complete spectral trace measured on a diffraction-limited spot over a time window of 400\,s is plotted in Fig.\,\ref{fig:SI_spectral_diffusion}. Almost all resolvable peaks can be assigned to individual molecules by comparing with the vibronic fingerprint of perylene. We identify in total six distinct molecular emission traces, denoted as Molecule 1 - Molecule 6, and corresponding digits in Fig.\,\ref{fig:SI_spectral_diffusion}(b). Limited by the sensitivity of the spectrometer, only the ZPL and up to two vibronic peaks could be reliably identified for Molecule 1 - Molecule 4. In contrast, Molecule 5 and Molecule 6 exhibit a more complete vibronic fingerprint.

The traces of all identifiable ZPL and vibronic peaks are presented in Fig.\,\ref{fig:SI_spectral_diffusion}(b). For each of the molecules, the ZPL and its vibronic peaks follow the same spectral diffusion trajectory, indicating that the full vibronic fingerprint shifts synchronously. The synchronized drift further reinforces the assignment of these peaks to the same single molecule and suggests that the dominant perturbation primarily shifts the entire electronic transition, while leaving the internal energy spacing of the associated vibronic states unchanged.

\begin{figure}[h!]
    \centering
    \includegraphics[width=0.8\linewidth,trim=0cm 0.5cm 0cm 1.7cm,clip]{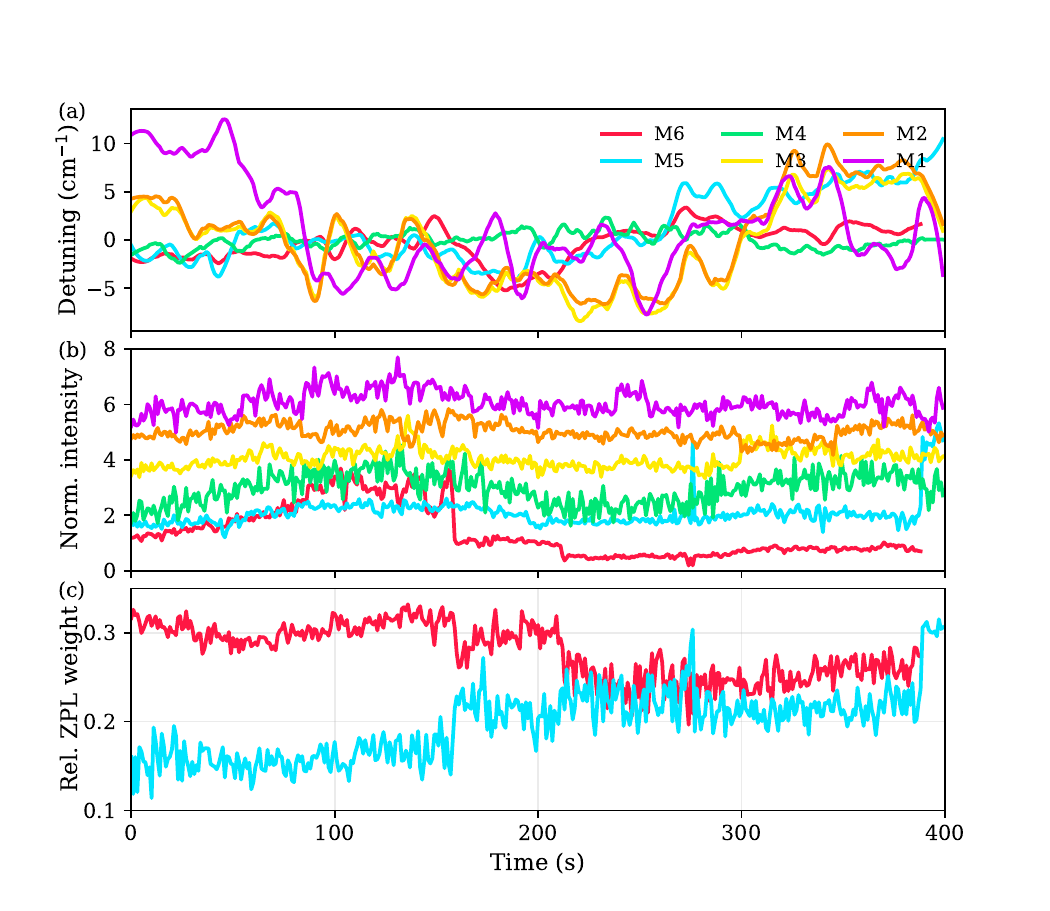}
    \caption{Extracted three time-dependent observables for the six molecules. (a) Spectral diffusion trajectories of the six molecules. (b) Median-normalized ZPL peak heights of molecules, vertically offset by constant amounts for clarity. (c) Defined relative ZPL weight for Molecule 5 and Molecule 6.}
    \label{fig:SI_spec_dif_obs}
\end{figure}

To gain further insights, we extract three time-dependent observables for each molecule: the spectral diffusion trajectory $ \Delta \nu(t) $, the ZPL peak intensity $I_\mathrm{ZPL}(t)$, and a relative ZPL weight $w_\mathrm{ZPL}(t)$ defined with respect to a fixed set of vibronic peaks defined as $w_\mathrm{ZPL}(t)=A_{\mathrm{ZPL}}/(A_{\mathrm{ZPL}}+A_{\mathrm{6fv}})$. Here, $A_{\mathrm{ZPL}}$  is the integral over a $10~\mathrm{cm}^{-1}$ width window centered on the ZPL position, and $A_{\mathrm{6fv}}$ is the sum of the integrated spectral window based on the six fundamental vibrational modes plotted in Fig.\,2(d) of the main manuscript.  The resulting time traces for the three observables are presented in Fig.\,\ref{fig:SI_spec_dif_obs}.

Fig.\,\ref{fig:SI_spec_dif_obs}(a) shows the diffusion temporal trajectories. Among the six molecules, only Molecule 2 and Molecule 3 follow the same spectral diffusion pattern. The remaining molecules do not show obvious mutual correlation in their frequency trajectories. In Fig.\,\ref{fig:SI_spec_dif_obs}(b), the relative variations of ZPL emission intensity $I_\mathrm{ZPL}(t)$ are presented after normalization to their median value, and applying a constant vertical offset for better visualization. The ZPL intensity remains relatively stable for Molecule 1-Molecule 5 around their respective baselines. By contrast, Molecule 6 alone shows pronounced intensity variation, indicating a molecule-specific process rather than a global effect across the diffraction-limited spot. A reasonable interpretation is that Molecule 6 undergoes a local environment change, possibly local charge-state fluctuations\cite{SI-Ciancico-2025-ChargeDefects}, that modifies its emission efficiency. Furthermore, in the later stage, the intensity of Molecule 6 completely vanishes, while Molecule 5 shows a strong increase at the same time. This anti-correlation appears to be specific to the Molecule 5-Molecule 6 pair and may indicate some degree of mutual coupling. However, whether it reflects an energy-transfer-like process cannot be determined confidently by the present data.

As shown in Fig.\,\ref{fig:SI_spec_dif_obs}(c), the relative ZPL weight $w_\mathrm{ZPL}(t)$ of Molecule 5 and Molecule 6 exhibits stage-like behavior. Molecule 5 shows a three-step evolution from a lower-$w_\mathrm{ZPL}$ to a higher-$w_\mathrm{ZPL}$ regime. Molecule 6, in contrast, starts from a relatively high $w_\mathrm{ZPL}$ level, undergoes a clear but transient reduction at intermediate times, and ends at a further lower $w_\mathrm{ZPL}$ stage. Such behavior suggests that these two molecules may possess a small number of metastable local configurations associated with changes in the ZPL-vibronic branching ratio.\cite{kulzer1997single,bordat2000elucidation}. It should be noted that $w_\mathrm{ZPL}(t)$ is not fully independent of the peak height $I_\mathrm{ZPL}(t)$, because $I_\mathrm{ZPL}(t)$ significantly contributes to $A_{\mathrm{ZPL}}$. Accordingly, an increase in $I_\mathrm{ZPL}(t)$ can raise $w_\mathrm{ZPL}(t)$. However, this mathematical effect does not fully account for the behavior observed in Molecules 5 and 6. For Molecule 5, the increase in $w_\mathrm{ZPL}(t)$ occurs while $I_\mathrm{ZPL}(t)$ remains relatively stable or even slightly decreases over the same period. This implies that the rise of $w_\mathrm{ZPL}(t)$ in Molecule 5 points to a suppression of the vibronic peak contribution. For Molecule 6, the variations in $I_\mathrm{ZPL}(t)$, including both its gradual increase and two rapid drops, do not lead to a synchronous change in $w_\mathrm{ZPL}(t)$. This suggests that the large intensity modulation in Molecule 6 is not directly tied to a long-term redistribution of the relative ZPL spectral weight.

Overall, although the present data do not uniquely identify the microscopic origin of the spectral evolution, they support a picture in which the observed emission dynamics arise from both continuous molecule-specific local perturbations and occasional switching between a small number of metastable local configurations.

\section{Sample Annealing}

In this section, we discuss the influence of high-temperature baking on the spectral stability. Here, we used Sample 2 in which perylene molecules had already been encapsulated in the hBN stack and baked it in air at $750\,^{\circ}\mathrm{C}$ for 4\,h. We expected this process to remove small trapped molecules, such as oxygen and water, from the stack even after encapsulation, and thereby potentially improve the spectral stability of perylene.

\begin{figure}[h!]
    \centering
    \includegraphics[width=1\linewidth,trim=0cm 0cm 0cm 0cm,clip]{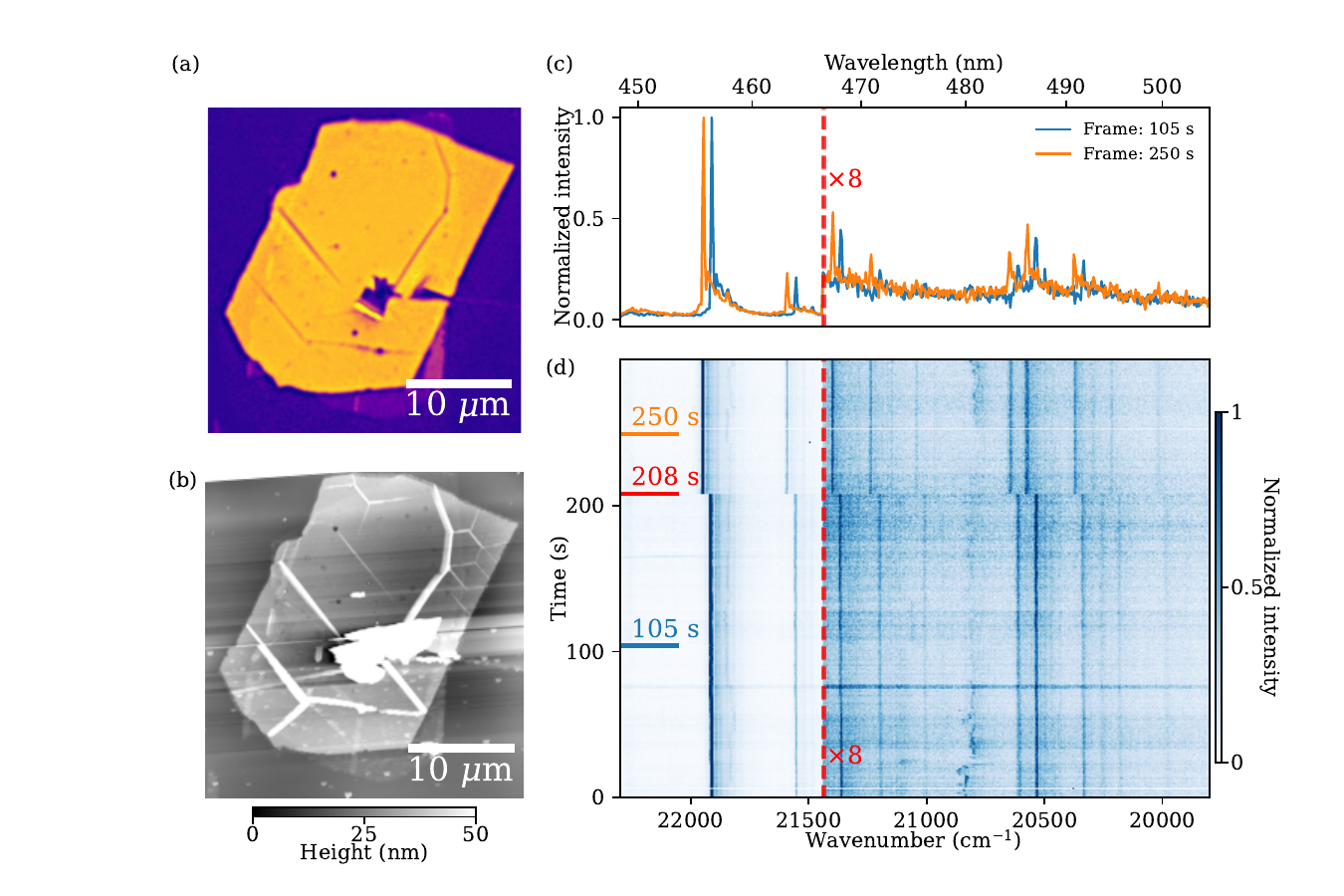}
    \caption{(a), (b) Overview optical microscope and AFM images of annealed sample 2. (c) Representative spectra extracted from the spectral evolution before and after a frequency drift event, corresponding to the horizontal markers in (d). (d) Spectral evolution of one perylene emission site on the annealed sample. The emission remains spectrally stable over several minutes except for a frequency drift at 208\,s, marked by the red horizontal line.}
    \label{fig:SI_annealed}
\end{figure}

The optical microscope and AFM images after annealing are shown in Figs.\,\ref{fig:SI_annealed}(a),(b). Compared with the original images before annealing shown in Fig.\,\ref{fig:SI_samples_overview}, the two hBN layers can still be resolved, indicating that the overall flake geometry is preserved after annealing. However, additional cracks and dot-like features appear within the hBN layers. These structural changes may originate from thermally induced stress, possibly related to the mismatch in thermal expansion coefficients between hBN and the silicon substrate. In addition, a region with strong optical contrast appears within the hBN flake. In the corresponding AFM image, this area shows a pronounced topographic contrast, indicating local damage or deformation of the hBN layer.

Optical measurements suggest that annealing improves the spectral stability of perylene. As shown in Fig.\,3(c) of the manuscript and Fig.\,\ref{fig:SI_annealed}(d), on the annealed sample, the emission traces of perylene molecules are much more stable, with only occasional drifts in frequency and intensity over several minutes. This is a proof that removing small molecules from the stack significantly removes the noise seen by perylene molecules, and these noise sources could be due to bond-flips or change of charge states in these unwanted molecules. Under efficient resonant excitation, stable emission remains for up to around 20\,minutes. At later times, they may exhibit photobleaching.

\section{Hyperspectral imaging}

To evaluate the spatial distribution of perylene molecules, we performed hyperspectral imaging to acquire emission spectra over a $10\,\mu\mathrm{m} \times 10\,\mu\mathrm{m}$ region of interest (ROI). In this measurement, the excitation laser is focused onto a diffraction-limited spot on the sample. The scattered light from the focal point is optically filtered using long-pass filters to block the excitation laser. Then, the remaining fluorescence signal passes through a confocal pinhole to suppress the background fluorescence and stray light. The signal is subsequently coupled into the spectrometer. After recording the spectrum at one spot, the laser focus is scanned across the entire ROI by using a steering mirror integrated in a 4f-telecentric imaging system, as shown in Fig.\,\ref{fig:SI_expsetup}.

A fluorescence intensity map can be generated by integrating the spectral intensity at each scanned position, as shown in Figs.\,5(c),(g) of the main manuscript. However, not all the strong fluorescence observed in our samples originates from perylene.  {For example, }broad emission bands at around 570\,nm and 620\,nm are also detected in our sample [see the left flank of black curves in Figs.\,\ref{fig:SI_LNCC}(a–d)]. These two bands match the spectrum of defect-related single photon emitters in hBN\,\cite{xu2018singlephoton,zeng2024single}. Therefore, instead of relying solely on fluorescence intensity, we identify perylene emission sites with spectral features across the hyperspectral map matched to a reference vibronic fingerprint shown in Fig.\,2(c) of the manuscript.

For the datasets recorded under 450\,nm excitation, the identification is performed using the local normalized cross-correlation (LNCC)\cite{lewis1995fast} method, i.e., a sliding-window zero-mean normalized cross-correlation. LNCC matching considers not only the relative spacing between vibronic peaks but also the overall spectral envelope, which improves the matching confidence for spectra with clean perylene fingerprints. By contrast, for the measurements conducted under 423\,nm excitation, the matching is only based on the spacing between vibronic peaks. The higher photon energy of 423\,nm lasers is more likely to excite additional background fluorescence, possibly from organic residues or other unidentified species in the sample. These background contributions may introduce broad spectral features within the perylene emission window and distort the profile-based LNCC matching result. Thus, peak-spacing matching is used to identify the fingerprints mainly through the relative spacing between dominant emission peaks of perylene, making the analysis less sensitive to variations in the complex spectral background.

\subsection{LNCC matching}

To do the LNCC matching, the spectra are converted to the wavenumber domain before pattern matching is conducted. Besides, both the reference pattern and the target spectrum are interpolated onto an evenly spaced wavenumber grid.

\begin{figure}[h!]
    \centering
    \includegraphics[width=1\linewidth]{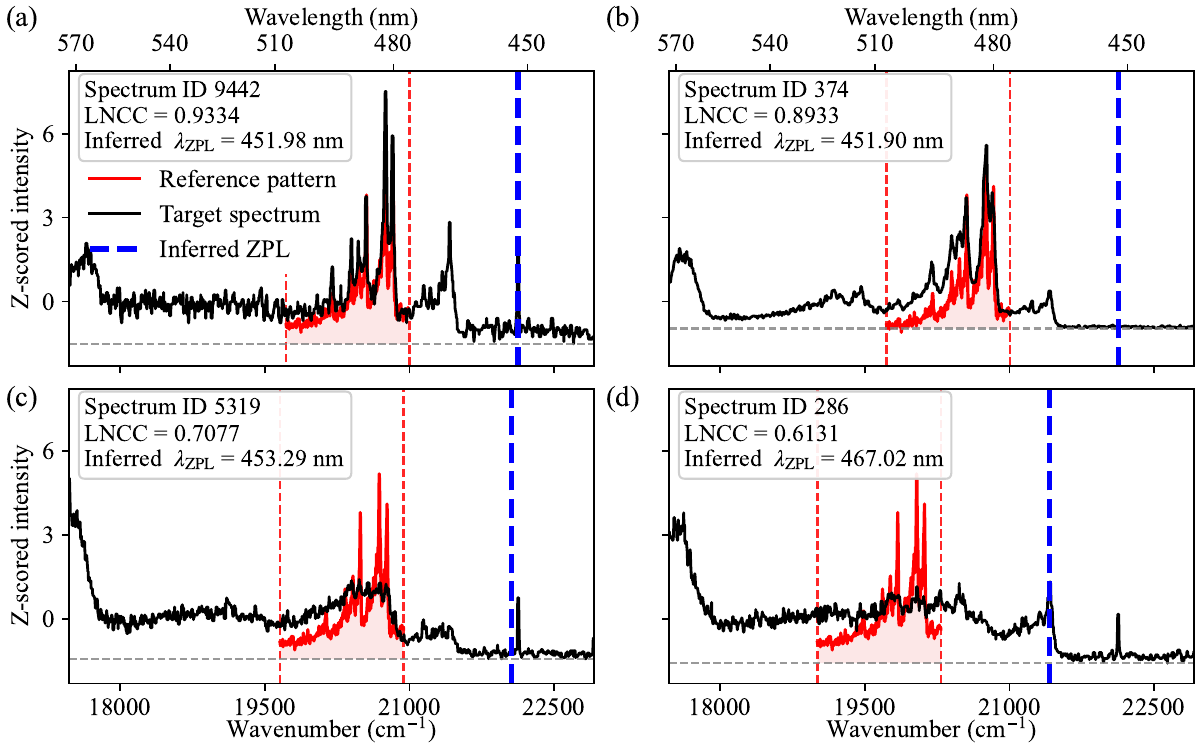}
    \caption{Representative examples of LNCC matching of perylene spectra. In all panels, the black and red curves denote the target spectrum and the reference pattern, respectively, while the blue dashed line marks the inferred ZPL position. (a) A high-confidence case with narrow spectral features and clear correspondence to the reference pattern. (b) A broader spectrum case. The overall lineshape and peak positions remain consistent with the reference, supporting reliable ZPL inference. (c) A case where peak positions are assigned, but the overall profile fails to match; therefore, it is regarded as unreliable. (d) A low-LNCC case, where no clear spectral correspondence can be identified, indicating an invalid match.}
    \label{fig:SI_LNCC}
\end{figure}

The reference template window, shown by the red shaded area in Figs.\,\ref{fig:SI_LNCC}(a–d), is denoted by \(x_i\), with \(i=1,2,\ldots,N\), and the target spectrum is written as \(y_j\), with \(j=1,2,\ldots,M\), where \(M>N\). The reference window then slides across the target spectrum, shown by the black curves in Figs.\,\ref{fig:SI_LNCC}(a-d), to evaluate the correlation at different lag positions. For each lag position \(k\), the corresponding target window is defined as \(y_i^{(k)} = y_{k+i-1}\), with \(i=1,2,\ldots,N\). Here, we apply $Z$-score normalization to each local target window rather than to the entire target spectrum. The corresponding $Z$-scores of the reference window and the normalized target window at lag \(k\) are written as
$Z_x(i) = ({x_i-\mu_x})/{s_x}$, and $
Z_y^{(k)}(i) = ({y_i^{(k)}-\mu_k})/{s_k}$, with $i=1,2,\ldots,N$, where \(\mu_x\) and \(s_x\) are the mean and standard deviation of the reference window, while \(\mu_k\) and \(s_k\) are the mean and standard deviation of the \(k\)–th target window. The correlation coefficient at lag position \(k\) is then calculated as
\begin{equation}
\mathrm{LNCC}(k)=\frac{1}{N}\sum_{i=1}^{N} Z_x(i)\,Z_y^{(k)}(i).
\end{equation}

Compared with a global normalization of the entire spectrum, LNCC suppresses the influence of absolute scale differences and reduces the risk that a few strong coincident features or slow baseline variations dominate the general trend\,\cite{avants2008symmetric}. The maximum LNCC value over all lag positions is taken as the matching score for each spatial position and then used to construct correlation maps shown in Figs.\,5(d),(h) of the main manuscript. 

Four representative evaluation results using LNCC are shown in Figs.\,\ref{fig:SI_LNCC}(a–d). In many cases of the emission spectra, the excitation wavelength lies close to the emitters' ZPL, resulting in the ZPL transition being filtered out by the longpass filter. However, when a reliable vibronic fingerprint match is obtained, the ZPL position can still be reconstructed from the known spectral spacing between the ZPL and vibronic peaks, as shown by the dashed blue lines in Figs.\,\ref{fig:SI_LNCC}(a–d).

\subsection{Peak-spacing matching}

\begin{figure}[h!]
    \centering
    \includegraphics[width=1\linewidth,trim=0cm 0cm 0cm 0cm,clip]{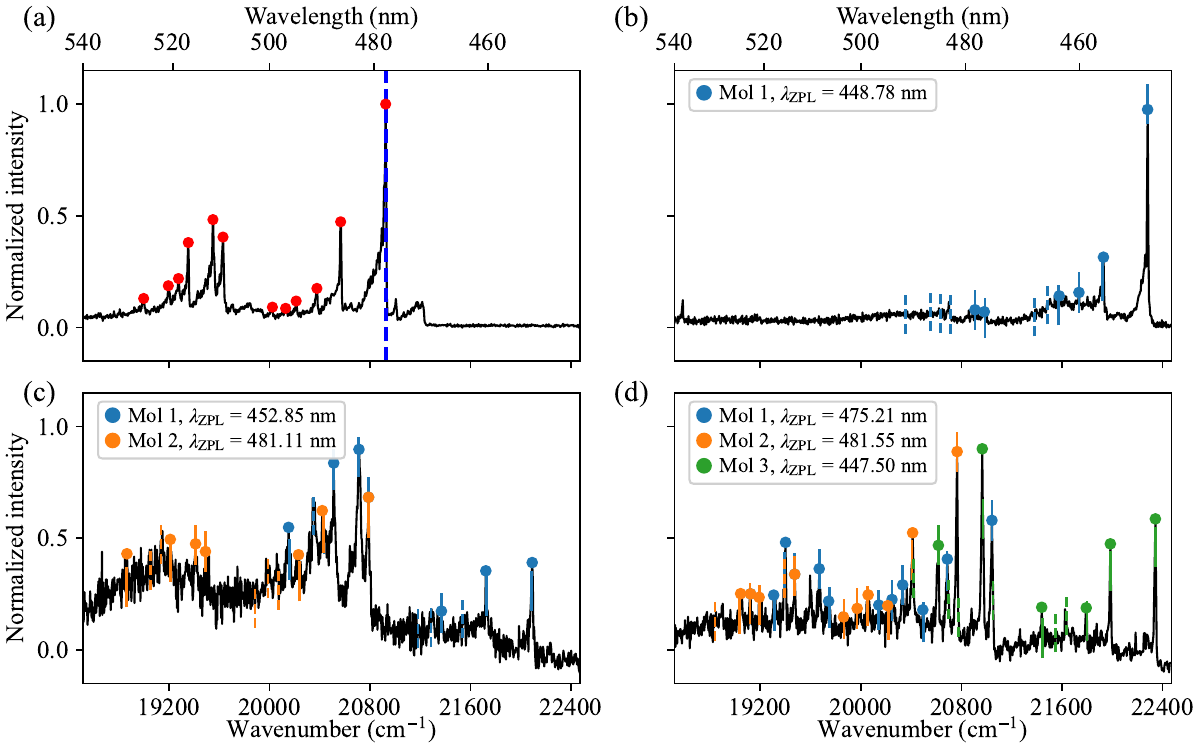}
    \caption{The reference spectrum and representative examples of peak-spacing matching of perylene spectra. In all panels, the black curves denote the measured spectra. (a) Reference perylene spectrum used to define the vibronic fingerprint. The red markers indicate the characteristic peaks used for the peak-spacing analysis, while the blue dashed line marks the ZPL position. (b–d) Representative target spectra analyzed using the peak-spacing method. Colored markers denote peaks assigned to different perylene molecules, and the corresponding inferred ZPL positions are listed in the legends. Dots and solid lines indicate the identified transition peaks, while dashed lines mark the expected peak positions based on the reference pattern.}
    \label{fig:SI_peak_space}
\end{figure}

For the datasets measured under 423\,nm excitation, perylene emission sites are identified using a peak-spacing-based fingerprint matching method. After the spectra are converted to the wavenumber domain and interpolated onto an evenly spaced wavenumber grid, the characteristic peaks of the reference spectrum are extracted, which are marked with red dots in Fig.\,\ref{fig:SI_peak_space}(a). These peaks are then used to define the relative peak spacing. For each target spectrum, a smoothly varying baseline is estimated from the background fluorescence and subtracted, followed by peak detection based on the local prominence of the baseline-corrected spectra.

Candidate perylene fingerprints are then generated by assigning target peaks as possible characteristic peaks and comparing the expected fingerprint peak positions with the measured data. A candidate is accepted only when a sufficient number of reference fingerprint features are found within the predefined wavenumber tolerance, together with additional constraints on the relative prominence of the dominant peaks. Since multiple perylene fingerprints may coexist within the same spectrum, candidates sharing too many of the same target peaks with others are removed to avoid double counting. The remaining candidates are assigned as possible individual perylene emission sites.

Representative matching results obtained using this peak-spacing method are shown in Figs.\,\ref{fig:SI_peak_space}(b-d), where peaks assigned to different perylene molecules are plotted in different colors. Furthermore, corresponding ZPL wavelength of each recognized molecule can also be reliably inferred. Compared with the spectra shown in Fig.\,\ref{fig:SI_LNCC}, these spectra recorded under 423\,nm excitation may suffer from reduced photon collection efficiency at the detection window from approximately 440\,nm to 460\,nm, due to chromatic aberration. Nevertheless, multiple perylene molecules can still be clearly recognized from their characteristic fingerprints, even when the overall spectral envelope is distorted.

\begin{figure}[h!]
    \centering
    \includegraphics[width=0.95\linewidth,trim=0.65cm 0.2cm 0.45cm 0.2cm,clip]{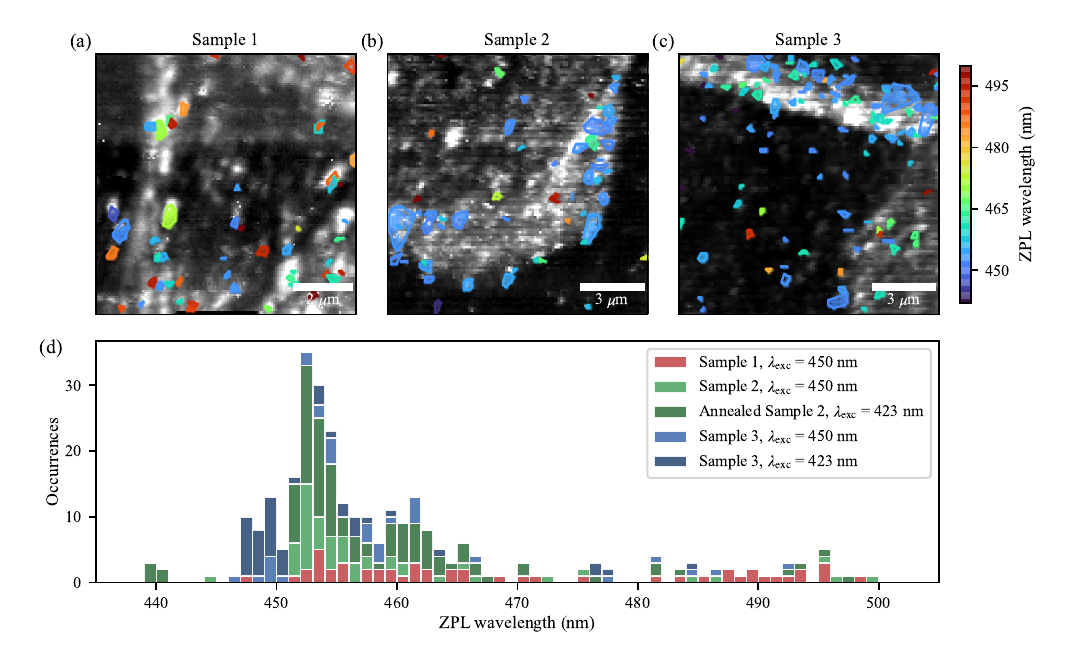}
    \caption{Spatial distribution of emission sites and corresponding histograms of inferred ZPL wavelength across the three samples. (a–c) Overlay maps of emission ensembles on hyperspectral maps of Samples 1--3. The color of the ensembles reflects the inferred ZPL wavelength. (d) Summary histogram of perylene ZPL wavelength across samples and excitation wavelengths.}
    \label{fig:SI_ZPL_distribution}
\end{figure}

Then, emission sites with inferred ZPL positions can be spatially mapped across the hyperspectral scan. Spectral matching may still produce isolated false-positive pixels due to noise or accidental partial similarity to the reference pattern. Therefore, the identification of valid emission sites is based not only on spectral matching but also on spatial continuity.

For each spatial position with a valid spectrum, we consider its planar coordinates \(\mathbf{r}_i=(x_i,y_i)\) and inferred ZPL wavelength \(\lambda_i\). A pixel is first retained as a candidate pixel only when it satisfies specific LNCC thresholds for LNCC-based matching or exhibits sufficient perylene fingerprint features in the peak-spacing-based matching. The retained candidate pixels are then grouped using a graph-based grouping rule. Two candidate pixels \(i\) and \(j\) are connected when they satisfy both a spatial proximity condition, $\|\mathbf{r}_i-\mathbf{r}_j\|\le d_0$, and a spectral consistency condition, $|\lambda_i-\lambda_j|\le \Delta\lambda_0$, where \(d_0\) is the maximum spatial linking distance and \(\Delta\lambda_0\) is the tolerance for ZPL position. The connected components of this graph are then taken as preliminary emission ensembles.

To recover edge pixels that likely belong to the same emission origin, a single-pass extension step is further applied. For a specific ensemble \(E\), a candidate pixel \(j\) identified by a KD-tree-based search\,\cite{friedman1977algorithm} can be added to ensemble \(E\) if it is located within the extension radius and ZPL wavelength tolerance. 

After grouping and extension, ensembles containing fewer than three valid candidate pixels are neglected because they are more likely to originate from noise than from stable emission. The retained ensembles are then overlaid on the hyperspectral fluorescence map, with the color indicating their corresponding ZPL wavelengths, together with the overall histogram of the ZPL wavelength throughout the sample shown in Fig.\,\ref{fig:SI_ZPL_distribution}. The spatial maps reveal substantial differences in the distribution of emission sites. In the pristine stacking region, the emission ensembles are generally smaller and spatially confined. In contrast, in the morphologically irregular region, emission is distributed over extended areas, where multiple sites can exhibit similar spectral features, hinting at similar local environments within each extended region. In addition, the extracted ZPL wavelengths across all emission ensembles span a broad range from 440\,nm to 500\,nm, as shown in Fig.\,\ref{fig:SI_ZPL_distribution}(d), indicating heterogeneous local insertion configurations in the sample.

A comparison of the spectral performance of perylene in different matrices, as well as terrylene adsorbed on an hBN surface, is shown in Tab.\,\ref{tab:SI_perylene_in_different_matrices}. We observe that the linewidth for our measurement is significantly higher than that for perylene in organic matrices. We attribute this to a possibly higher pure dephasing rate of perylene embedded in hBN. This also would explain why we did not observe power broadening, as dephasing would be the dominant broadening mechanism here.

\begin{table}[h!]
\centering
\small
\setlength{\tabcolsep}{4pt}
\begin{tabular}{ccccc}
\hline
Guest &
Host &
Linewidth (MHz) &
$\lambda_{ZPL}$ (nm) \\
\hline
Perylene & encapsulated in hBN & $\approx$ 2800 & $\approx$ 440 - 500\\
Perylene & Polyethylene & $\approx$ 100  & $\approx$ 450\\
Perylene-d12 & Dibenzothiophene & $\approx$ 58 & $\approx$ 454 \\
Perylene & Anthracene & no single molecule observation & $\approx$ 449 \\
Perylene & Biphenyl & $\approx$ 140 & $\approx$ 445 \\
Perylene & n-Nonane & $\approx$ 34 & $\approx$ 444 \\
Perylene & o-Dichlorobenzene & $\approx$ 53 & $\approx$ 447 \\
Terrylene & on hBN surface & $\approx$ 450 & $\approx$ 582 \\
\hline
\end{tabular}
%\end{adjustbox}
\caption{Comparison of perylene (and terrylene at the bottom) as guest single molecule in different host systems. Except for our own measurements of perylene encapsulated in hBN and perylene in anthracene, the information for all the other displayed guest-host systems was taken from Adhikari et al.\cite{Orrit_2024}. For perylene in anthracene, there is no single molecule information available.}
\label{tab:SI_perylene_in_different_matrices}
\end{table}
\newpage

\CombinedSIBibliography

\end{document}